\documentclass[11pt,a4paper]{article}
\usepackage[a4paper,margin=2.5cm]{geometry}
\usepackage[T1]{fontenc}
\usepackage{graphicx}
\usepackage{amsmath,amsfonts,amssymb}
\usepackage{url}
\def\lt{<}

\newcommand{\mcWidth}[1]{\csname TY@F#1\endcsname+\tabcolsep}

\def\processVert{\ifmmode|\else\textbar\fi}

\let\arxivgraphicx\includegraphics
\newcommand{\arxivfitbox}[2]{%
  \ifdim#1>\linewidth
    \resizebox{\linewidth}{!}{\usebox0}%
  \else\ifdim#2>0.61\textheight
    \resizebox{!}{0.61\textheight}{\usebox0}%
  \else\usebox0\fi\fi}
\renewcommand{\includegraphics}[2][]{%
  \sbox0{\arxivgraphicx[#1]{#2}}%
  \arxivfitbox{\wd0}{\ht0}}
\makeatletter
\long\def\@makecaption#1#2{%
  \vskip\abovecaptionskip
  {\footnotesize\textbf{#1.} #2\par}}
\makeatother

\title{Femtosecond Free-Space Time-Frequency Transfer at km/s Relative Motion with Doppler Compensation}

\date{}
\author{\relax}

\begin{document}

\maketitle

\begin{center}
\large
Jian-Yu Guan\textsuperscript{1,2,*}, Qi Shen\textsuperscript{1,2,3,*}, Lei Hou\textsuperscript{1,2,*}, Jin-Jian Han\textsuperscript{1,2,3}, Shen Zhang\textsuperscript{1,2,3}, Feng-Heng Che\textsuperscript{1,2,3}, Yi-Cheng Zhang \textsuperscript{1,2,3}, Shi Chen\textsuperscript{1,2,3}, Yao-Dong Hu\textsuperscript{1,2,3}, Zhi-Kang Qian\textsuperscript{1,2,3}, Yong Wang\textsuperscript{5}, Min Li\textsuperscript{1,2}, Shao-Mao Wang\textsuperscript{1,2}, Dan-Yang Zhu\textsuperscript{1,2}, Fa-Xi Chen\textsuperscript{1,4}, Li-Bo Li\textsuperscript{4}, Esamdin Ali\textsuperscript{5}, Ji-Gang Ren\textsuperscript{1,2,3}, Juan Yin\textsuperscript{1,2,3}, Rong Shu\textsuperscript{1,6}, Cheng-Zhi Peng\textsuperscript{1,2,3}, Hai-Feng Jiang\textsuperscript{1,2,3}, Qiang Zhang\textsuperscript{1,2,3,4}, Jian-Wei Pan\textsuperscript{1,2,3}
\end{center}

\begin{center}
\footnotesize
$^{1}$Hefei National Laboratory, University of Science and Technology of China, Hefei, China \\
$^{2}$Shanghai Research Center for Quantum Sciences and CAS Center for Excellence in Quantum Information and Quantum Physics, University of Science and Technology of China, Shanghai, China \\
$^{3}$Hefei National Research Center for Physical Sciences at the Microscale and School of Physical Sciences, University of Science and Technology of China, Hefei, China \\
$^{4}$Jinan Institute of Quantum Technology and CAS Center for Excellence in Quantum Information and Quantum Physics, University of Science and Technology of China, Jinan 250101, China \\
$^{5}$Xinjiang Astronomical Observatory, Chinese Academy of Sciences, Urumqi 830011, China \\
$^{6}$Innovation Academy for Microsatellites of CAS, China\\
$^{\ast}$These authors contributed equally to this work.
\end{center}

\vspace{1em}

\begin{abstract}
A global optical clock network holds great promise for a wide range of applications, including international metrology--particularly the redefinition of the ``second''~\cite{cgpm_2022_resolution_5, dimarcq2024roadmap, lindvall2025coordinated}--as well as navigation, geoscience~\cite{mcgrew2018atomic, grotti2018geodesy, takamoto2020test, bothwell2022resolving, lisdat2016clock}, and fundamental physics~\cite{collaboration2021frequency, tsai2023direct, kennedy2020precision, filzinger2023improved}. Satellite-based high-precision time-frequency dissemination is essential for realizing such a network. Despite remarkable progress in ground-based free-space simulations over high-loss channels~\cite{shen2022free, caldwell2023quantum}, the Doppler effect arising from satellite motion has remained an unsolved barrier. Here we demonstrate femtosecond two-way time-frequency transfer over a 113-km free-space link under simulated Doppler shifts of up to 1.55\,km/s. To overcome the Doppler barrier, we first reproduce satellite-like Doppler shifts of up to 1.55\,km/s on the ground by locking a frequency comb to a broadly tunable laser. We then correct the resulting disturbances with two complementary techniques: a dual-channel sampling architecture that compensates for optical dispersion and radio frequency (RF) nonlinearity while eliminating Nyquist dead zones, and an iterative interferogram reconstruction algorithm that removes Doppler-induced distortion. After correction, residual systematic errors are within \(\pm 10\)~fs over the full 1.55~km/s range. Using simulated medium earth orbit (MEO) and geosynchronous orbit (GSO) satellite trajectories, we achieve fractional frequency instability of $3.9\times 10^{-19}$ at 4,600\,s and $3.6\times 10^{-19}$ at 23,600\,s, respectively. This extends the operational velocity range by nearly two orders of magnitude beyond previous free-space demonstrations~\cite{bergeron2019femtosecond,sinclair2019femtosecond}, removing the last major technical barrier to a global femtosecond-precision satellite-based time-frequency link.
\end{abstract}

\section{Introduction}

State-of-the-art optical clocks~\cite{aeppli2024clock, marshall2025high, arnold2026lu+, jia2026improved, zhang2026liquid} have achieved fractional frequency uncertainties below $1\times 10^{-18}$, opening up unprecedented opportunities for fundamental science and metrology via a future global clock network. However, the critical bottleneck in realizing a global-scale optical clock network lies in the capability to compare these standards across intercontinental distances. Although bidirectional optical fiber links can effectively interconnect regional clock networks~\cite{predehl2012optical, droste2013optical, schioppo2022comparing, chen202610}, their fundamental incompatibility with commercial unidirectional transoceanic submerged cables precludes intercontinental comparisons. Optical two-way time-frequency transfer (O-TWTFT) based on dual-comb linear optical sampling (LOS) offers a highly scalable solution for satellite-ground links. By delivering femtosecond-level precision paired with a vast ambiguity-free range, this dual-comb approach enables distant ground stations to cancel link noise and evaluate remote clocks by extending the classic  satellite common-view architectures~\cite{allan1980accurate, defraigne2015cggtts, huang2016beidou}.
 
\begin{figure}
    \centering
    \includegraphics[height=0.685\textheight,keepaspectratio]{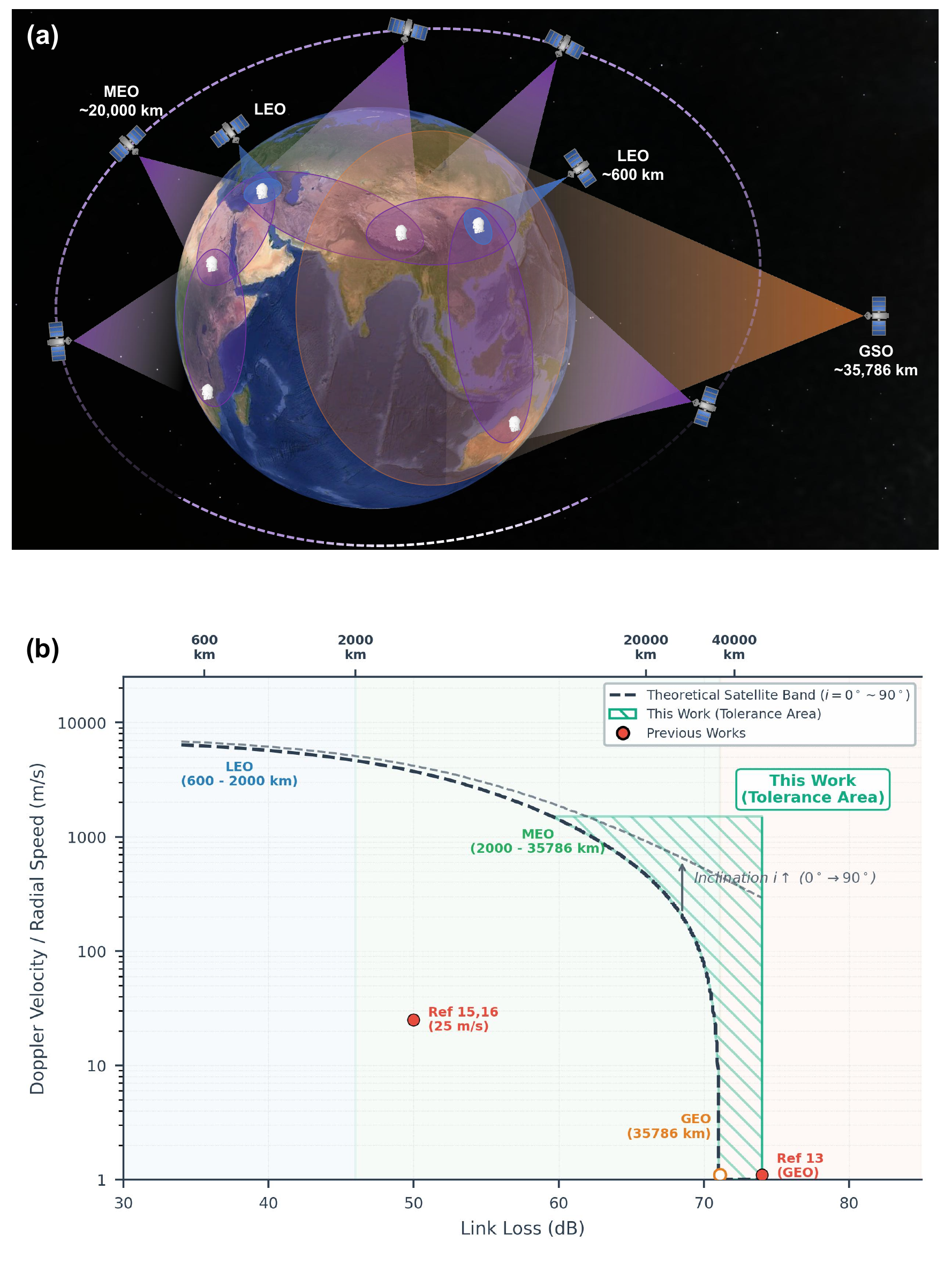}
    \caption{
    \textbf{Orbital geometries and the velocity-loss trade-off space for satellite-ground time-frequency transfer.}
    \textbf{a}, Schematic representation of satellite ground tracks and instantaneous visibility footprints across three primary orbital regimes: Low Earth Orbit (LEO, ~600 km), Medium Earth Orbit (MEO, ~20,000 km), and Geosynchronous Orbit (GSO, ~35,786 km). Ground-based optical terminals are represented by white domes. LEO constellations possess a narrow visibility footprint (blue cones) incapable of bridging intercontinental baselines. GSO project a massive hemispheric footprint (orange cone) supporting long-term, persistent common-view links, but their stationary regional confinement limits global network scaling. MEO satellites reconcile these requirements; their wide instantaneous footprints (purple cones) support daily intercontinental common-view windows of several hours while their dynamic ground tracks sweep the entire globe.
    \textbf{b}, Doppler velocity tolerance versus link loss capability across different orbital regimes. The black dashed curve represents the theoretical physical envelope for orbits ranging from LEO to GEO under varying orbital inclinations ($i = 0^\circ \text{ to } 90^\circ$). Shaded vertical regions indicate the typical parameter space for LEO (600-2,000 km, blue), MEO (2,000-35,786 km, green), and GSO/GEO (35,786 km, pink). The link attenuation is calculated assuming that both the transmitting and receiving telescopes have apertures of 1\,m, with detailed link parameters described by Shen et al.~\cite{shen2021experimental} (see its Supplementary Information). The green hatched region highlights the demonstrated operational domain of this work, showcasing simultaneous operability under severe link attenuation (up to 74 dB) and high-speed Doppler dynamics (up to 1.55 km/s). Red circles designate previous pioneering demonstrations, which were either constrained to low-velocity regimes at moderate loss (25 m/s under ~50 dB attenuation) or static ground-emulation configurations under high attenuation (GEO simulations with zero relative velocity). 
    }
    \label{fig:orbitAndFreShift}
\end{figure}

The suitable orbits for the global-scale optical clock network should provide a visibility coverage wide enough to span distinct continents while maintaining a global ground track that avoids regional confinement. As schematically illustrated in Fig.~\ref{fig:orbitAndFreShift}a\nocite{shen2021experimental}, Medium Earth Orbit (MEO) satellites reconcile these competing requirements, offering a vast coverage area that supports daily intercontinental common-view windows of several hours alongside global ground tracks that sweep the entire Earth. In contrast, while Geosynchronous Orbit (GSO) configurations can sustain long-term common-view links between two specific landmasses, their fixed hemispheric coverage restricts broader network scaling. Meanwhile, Low Earth Orbit (LEO) architectures cannot support direct intercontinental common-view links due to their limited instantaneous visibility footprints. Consequently, modern Global Navigation Satellite Systems (GNSS) predominantly leverage MEO and Geosynchronous Orbit (GSO) spacecraft situated at altitudes exceeding 20,000\,km~\cite{teunissen2017springer}. MEO constellations offer global scalability, while GSO configurations (including inclined orbits) provide continuous target-specific coverage. Both regimes thus constitute the foundational orbital resources for an intercontinental optical clock network.

Establishing these high-orbit satellite-ground links, however, requires concurrently overcoming two major physical barriers: extreme channel loss and massive motion-induced Doppler frequency shifts. Across different orbital configurations, these challenges exhibit an inherent trade-off (Fig.~\ref{fig:orbitAndFreShift}b): higher altitudes mitigate Doppler dynamics at the expense of severe geometric path loss, whereas lower orbits alleviate attenuation but introduce rapid velocity-dependent distortions. Geostationary Earth Orbit (GEO), a static subset of the GSO regime ($i=0^\circ$), represents a stationary shortcut characterized by maximum path loss but zero relative velocity. Yet, while static links have recently been emulated on the ground~\cite{shen2022free, caldwell2023quantum}, equatorial GEO orbital slots are scarce and commercially congested. Generalized global dissemination must therefore utilize inherently dynamic orbits such as MEO and inclined GSO tracks, which inevitably introduce line-of-sight radial velocities of up to 1\,km/s. Prior pioneering free-space experiments in the presence of motion~\cite{bergeron2019femtosecond, sinclair2019femtosecond} operated under channel losses of $\sim$50\,dB, but were restricted to modest velocities below 25\,m/s, leaving a substantial gap compared to
actual high-orbit speeds. Here, we report an experimental demonstration that simultaneously overcomes severe link attenuation and high-speed motion dynamics over a 113-km free-space link, completely encompassing both MEO and GSO operational regimes. 

\section{High-Speed Relative Motion Emulation}

Validating optical two-way transfer architectures for satellite-to-ground links strictly requires testing in the presence of kilometer-per-second platform motion. However, replicating physical motion at kilometer-per-second speeds on the ground over extended durations is practically intractable. Fortunately, high-precision acquisition, pointing and tracking (APT) frameworks have already been thoroughly matured in pioneering free-space quantum and laser communication missions~\cite{ren_ground--satellite_2017}. The remaining technical bottleneck, therefore, shifts from mimicking mechanical trajectories to actively mitigating the massive Doppler shifts imposed on the optical carrier. Consequently, a high-fidelity ground-based verification platform needs only to precisely synthesize the Doppler-induced frequency variations, obviating the need for physical acceleration or redundant tracking dynamics. For a single-frequency continuous-wave (CW) reference, the Doppler shift is expressed as $\Delta f_{\text{CW}} = f_{\text{CW}} \frac{v}{c}$, where $v$ represents the relative velocity and $c$ is the speed of light. This motion-free frequency translation is physically emulated here using a broadly tunable, ultra-stable laser stabilized via an offset-frequency Pound--Drever--Hall (PDH) locking scheme, allowing the output frequency to be continuously modulated by dynamically tuning the radio-frequency (RF) offset applied to an electro-optic modulator (EOM). A comprehensive description of the complete frequency-domain Doppler simulation architecture is provided in the Methods.

As noted previously, while single-frequency CW lasers are standard for optical frequency transfer over fiber links, O-TWTFT relies on optical frequency combs to bridge satellite-ground links. Dual-comb LOS ensures robust phase retrieval even during turbulence-induced link fades due to its high precision and vast ambiguity-free range. In a dual-comb architecture, the transmitted signal comprises thousands of phase-locked lines expressed as $f_n = n f_{\text{r}} + f_{\text{CEO}}$, where $n$ is the comb-tooth index, $f_{\text{r}}$ is the repetition rate, and $f_{\text{CEO}}$ is the carrier-envelope offset frequency. Under physical relative motion, the optical Doppler effect imparts a frequency shift strictly proportional to each optical carrier: $\Delta f_n = f_n \frac{v}{c} = n f_{\text{r}}\frac{v}{c} + f_{\text{CEO}}\frac{v}{c}$. Consequently, both $f_{\text{r}}$ and $f_{\text{CEO}}$ scale by the exact same ratio, manifesting as a proportional stretching of the entire comb spectrum rather than a rigid translation. Conventional acousto-optic or electro-optic modulators only apply a uniform frequency offset across all teeth (shifting $f_{\text{CEO}}$ alone) and thus fail to replicate this physical stretching. To overcome this limitation, we employ a self-referenced locking scheme in which both $f_{\text{CEO}}$ and the repetition rate are synchronized to a CW reference laser ($f_n \propto f_{\text{CW}}$). By dynamically tuning $f_{\text{CW}}$ via offset-locking, the optical comb is driven into an authentic Doppler breathing mode across the entire spectrum. This architecture synthesizes an optical carrier frequency excursion of up to 1\,GHz, equivalent to an operational radial velocity of 1.55\,km/s at 1550\,nm.

%For a unidirectional link (e.g., from Terminal B to Terminal A), this simulated Doppler shift is equivalent to a true relative velocity $v_{sim} = c \cdot f_{gap} / f_{CW}$. For the reverse path (Terminal A to Terminal B), the effect is slightly different because the local comb at Terminal B experiences a frequency shift. From the reference frame of Terminal B's sampling module, this is mathematically equivalent to the incoming signal comb from Terminal A experiencing an inverse frequency shift. The equivalent simulated velocity \(v_{sim}^{\prime }\) for this reverse path is therefore given by $v_{sim}' \approx -v_{sim}$ (for $v_{sim} \ll c$). This asymmetry only affects the sign of some compensation terms during clock error calculation. It does not compromise the validity of our high-speed motion simulation system. A detailed description of the complete frequency-domain Doppler simulation system is given in Methods. 

In this ground-based verification, modulating the optical frequency at Terminal B faithfully replicates the physical Doppler shift for the transmission from B to A ($v_{\text{sim}} = c \Delta f / f_{\text{CW}}$). For the reverse path (A to B), the unshifted signal from Terminal A beats against the shifted local comb at Terminal B. Within the linear optical sampling frame, this is mathematically equivalent to the incoming comb experiencing an opposite Doppler shift ($v_{\text{sim}}' \approx -v_{\text{sim}}$ for $v \ll c$). This reciprocal antisymmetry rigorously reproduces the Doppler-induced spectral dynamics in both link directions, with the only distinction from a moving space platform being an analytical sign inversion in the bidirectional non-reciprocal path correction ($Lv/2c^2$) during two-way clock-offset extraction (see Methods for detailed kinematics).

\section{Motion-Induced One-Way Errors and Compensation}

Accurate two-way clock-offset extraction fundamentally relies on unbiased one-way arrival-time measurements. While the kinematic path asymmetry ($Lv/2c^2$) is analytically corrected, motion-induced Doppler shifts severely corrupt the linear optical sampling (LOS) at each terminal. 

\begin{figure}
    \centering
    \includegraphics[height=0.618\textheight,keepaspectratio]{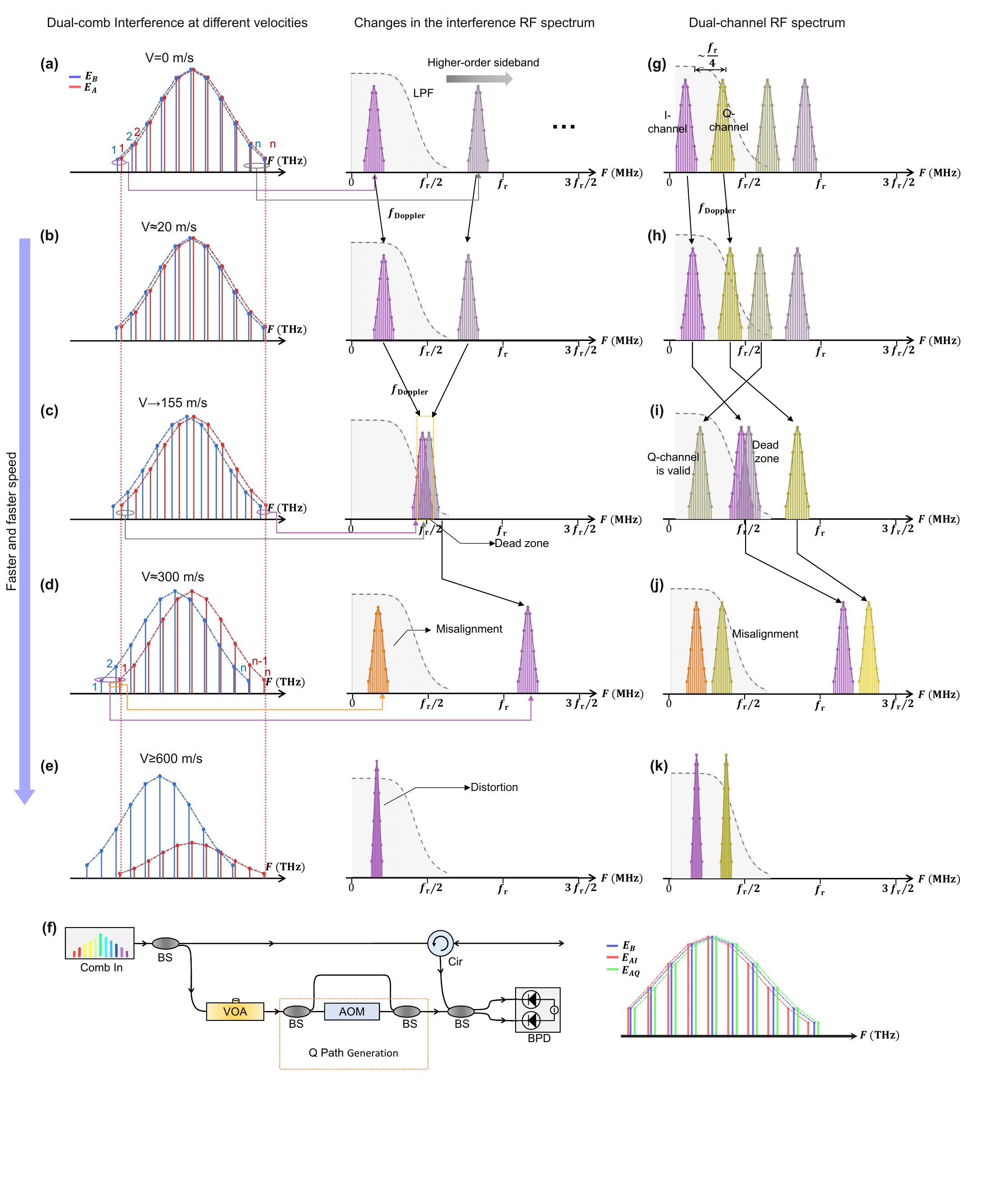}
    \caption{\scriptsize\textbf{Motion-induced impairments and dual-channel compensation scheme. a--e,} Impact of increasing Doppler velocity on single-channel dual-comb interferometry. Terminal~B is assumed to recede from Terminal~A at radial velocity $v$, decreasing the received signal frequency at A.
    \textbf{a,} Static baseline ($V = 0$\,m/s): The multi-heterodyne radio-frequency (RF) spectrum contains periodic harmonic bands. The fundamental baseband is isolated by a hardware low-pass filter (LPF) with a cutoff at $f_{\text{r}}/2$. The non-uniform electronic phase response across the detection chain is indicated by the curved dotted line.
    \textbf{b,} Sub-Nyquist regime ($V \approx 20$\,m/s): Carrier Doppler translation induces a rigid shift ($f_{\text{Doppler}}$) of the down-converted baseband spectrum across the curved electronic phase profile, breaking common-mode phase cancellation.
    \textbf{c,} Nyquist-folding boundary ($V \to 155$\,m/s): The shifted spectrum collides with the $f_{\text{r}}/2$ Nyquist boundary and overlaps with its folded alias, creating a measurement ``dead zone'' characterized by spectral ambiguity.
    \textbf{d,} Inter-comb-period regime ($V \approx 300$\,m/s): With the carrier Doppler shift traversing a full comb repetition rate ($f_{\text{Doppler}} \approx f_{\text{r}}$), down-conversion in the baseband originates from mismatched optical comb-tooth pairs (comb-tooth misalignment, e.g., tooth $n$ with $n-1$), introducing discrete, dispersion-dependent phase tilts.
    \textbf{e,} Hyper-dynamic envelope regime ($V \ge 600$\,m/s): Relativistic Doppler scaling of the repetition-rate offset alters $\Delta f_{\text{r}}$, inducing severe temporal envelope dilation/compression and invalidating conventional rigid-envelope matching.
    \textbf{f,} Optical schematic for the quadrant-shifted dual-channel architecture. The local oscillator (LO) comb is split into an unshifted Channel~I ($E_{\text{A,I}}$) and an AOM-frequency-shifted Channel~Q ($E_{\text{A,Q}}$) with a relative offset of $f_{\text{AOM}}$ before interfering with the signal comb ($E_{\text{B}}$) on a balanced photodetector (BPD). BS, beam splitter; VOA, variable optical attenuator; Cir, circulator.
    \textbf{g--k,} Dual-channel RF multi-heterodyne spectra corresponding to the operational regimes in \textbf{a--e}, demonstrating progressive error isolation and compensation.
    \textbf{g,} Static baseline ($V = 0$\,m/s): The Channel~I and Channel~Q spectra maintain an engineered quadrature separation of $\sim f_{\text{r,A}}/4$.
    \textbf{h,} Sub-Nyquist regime ($V \approx 20$\,m/s): Both channels shift synchronously. Because they sample distinct portions of the LPF passband, the differential arrival time $\tau_{\text{IQ}}(v)$ isolates the electronic RF phase nonlinearity.
    \textbf{i,} Nyquist-folding boundary ($V \to 155$\,m/s): The quarter-period spectral offset ensures mutually exclusive dead zones. When Channel~I encounters Nyquist folding at $f_{\text{r}}/2$, Channel~Q remains centered within the LPF passband, eliminating measurement dropouts.
    \textbf{j,} Inter-comb-period regime ($V \approx 300$\,m/s): In this regime ($f_{\text{Doppler}} \approx f_{\text{r}}$), down-conversion in both channels originates from mismatched comb-tooth pairs.
    \textbf{k,} Hyper-dynamic envelope regime ($V \ge 600$\,m/s): Both channels experience simultaneous envelope distortion where hardware interleaving reaches its physical limit; the iterative reconstruction algorithm is applied to restore the zero-velocity interferogram and extract the true one-way arrival time.
    }
    \label{fig:Interference}
\end{figure}

To establish a baseline for evaluating these distortions, we first consider dual-comb LOS under static conditions ($v = 0$, Fig.~\ref{fig:Interference}a), where the local oscillator (LO) comb A (repetition rate $f_{\text{r,A}}$) and the incoming signal comb B (repetition rate $f_{\text{r,B}}$) optically mix to down-convert the dense optical spectrum into radio-frequency (RF) beat notes. Because each signal comb line beats simultaneously with all teeth of the local comb, the resulting multi-heterodyne spectrum generates periodic RF beat notes repeating across intervals of $f_{\text{r,A}}/2$. Although each of these spectral bands inherently encodes the timing information, applying an analog low-pass filter (LPF) with a cutoff at $f_{\text{r,A}}/2$ is most conveniently chosen to isolate the baseband. This filtering retains exclusively the beats produced by the nearest comb-tooth pairings while rejecting higher-frequency replicas, yielding an unaliased interferogram. In this static regime, subtracting the phase of an initial unperturbed interferogram cancels out both the intrinsic optical chromatic dispersion and the electronic phase response, directly yielding the true one-way arrival time from the linear phase slope.

Prior pioneering experiments in the presence of motion~\cite{bergeron2019femtosecond, sinclair2019femtosecond} were restricted to modest velocities below 25\,m/s. Under such low dynamics (Fig.~\ref{fig:Interference}b), Doppler shifts alter the optical carrier by less than 16\,MHz and vary the repetition rate by only tens of hertz. Because the spectrum experiences only a minor rigid translation within the static filter passband, the teeth remain correctly paired. Under this narrow regime, the cross-ambiguity function (CAF) algorithm can numerically realign the spectra.

However, scaling to realistic satellite velocities approaching kilometer-per-second regimes shatters this low-speed approximation, triggering four cascading failure modes that are visually captured in Fig.~\ref{fig:Interference}b--e (see Methods for detailed mathematical derivations):
\begin{enumerate}
    \item \textbf{Sub-Nyquist regime ($V \approx 20$\,m/s, Fig.~\ref{fig:Interference}b):} The dominant impairment is the electronic RF phase nonlinearity. Carrier Doppler translation rigidly shifts the baseband spectrum across the detection chain. Because photodetectors and filters exhibit a non-uniform phase response (curved dotted line in Fig.~\ref{fig:Interference}a), shifting across this curved profile alters the effective group delay experienced by the interferogram, breaking common-mode phase cancellation and inducing hundreds of femtoseconds of motion-dependent timing errors.
    \item \textbf{Nyquist-folding boundary ($V \to 155$\,m/s, Fig.~\ref{fig:Interference}c):} The dominant threat shifts to periodic Nyquist dead zones. As the velocity approaches $\sim 155$\,m/s, the shifted spectrum collides with the $f_{\text{r,A}}/2$ Nyquist boundary and folds back onto itself. This spectral overlapping and aliasing destroys the phase structure, creating periodic dead zones (recurring every 155\,m/s) that cause complete measurement dropouts across an orbit.
    \item \textbf{Inter-comb-period regime ($V \approx 300$\,m/s, Fig.~\ref{fig:Interference}d):} Beyond the 155\,m/s threshold, a valid interferogram re-emerges. While the electronic RF phase nonlinearity persists across the shifted passband, a new, distinct systematic bias arises from the comb's intrinsic chromatic dispersion. Because the original baseband beat note has shifted out of the passband, the fixed filter now down-converts from a mismatched pair of optical comb teeth (exemplified in Fig.~\ref{fig:Interference}d, where tooth $n$ beats against tooth $n-1$). Owing to the comb's chromatic dispersion, this tooth-index slip adds an artificial phase tilt, imprinting a discrete, stepwise timing bias that accumulates to $\sim 100$\,fs across the velocity excursion.
    \item \textbf{Hyper-dynamic envelope regime ($V \ge 600$\,m/s, Fig.~\ref{fig:Interference}e):} The primary impairment escalates to motion-induced envelope distortion. At these extreme speeds, the Doppler scaling of the repetition rate alters the effective repetition-rate difference ($\Delta f_{\text{r}}$) by more than 20\%. This massive variation dramatically dilates or compresses the multi-heterodyne spectral envelope (visually captured by the broadened profile labeled ``Distortion'' in Fig.~\ref{fig:Interference}e), fundamentally invalidating the rigid-envelope assumption of conventional algorithms (such as CAF) and producing picosecond-level timing deviations.
\end{enumerate}

To systematically resolve these cascading failures, we implement a quadrant-shifted dual-channel linear optical sampling architecture (Fig.~\ref{fig:Interference}f). The local comb A is split into an unshifted I channel and a Q channel actively shifted by an AOM ($f_{\text{AOM}} \approx 47.26\text{ MHz} \neq f_{\text{r,A}}/4$), separated by a 1.5-ns optical delay. In the frequency domain, their down-converted spectra maintain a constant quadrature offset of approximately $f_{\text{r,A}}/4$ (Fig.~\ref{fig:Interference}g).

The principal innovation of this dual-channel configuration lies in its ability to quantitatively decouple and calibrate both the electronic RF nonlinearity and the optical comb dispersion without extra reference instruments. Across low and intermediate velocities, this interleaved architecture progressively resolves each failure mode:

\begin{itemize}
    \item In the sub-Nyquist regime ($V \approx 20$\,m/s, Fig.~\ref{fig:Interference}h), the differential arrival time between the two channels, $\tau_{\text{IQ}}(v) = \tau_{\text{Q}}(v) - \tau_{\text{I}}(v)$, acts as an in-situ diagnostic probe. Because the two channels sample different portions of the filter passband, the non-uniform electronic phase response maps into a continuous, periodic anti-symmetric curve centered at $f_{\text{AOM}}/2$, isolating the RF nonlinearity.
    \item At the Nyquist-folding boundary ($V \to 155$\,m/s, Fig.~\ref{fig:Interference}i), the quarter-period spectral offset ensures that the dead zones of the two channels are mutually exclusive. When Channel I falls into the Nyquist folding boundary, Channel Q remains comfortably centered within the clear LPF passband, seamlessly eliminating measurement dropouts.
    \item In the inter-comb-period regime ($V \approx 300$\,m/s, Fig.~\ref{fig:Interference}j), the carrier Doppler shift exceeds the comb repetition rate, driving the dual-channel spectra into comb-tooth mispairing regimes. Across a Doppler excursion, this tooth-index sliding imprints discrete, stepwise timing shifts onto $\tau_{\text{IQ}}$.
\end{itemize}

Because the continuous RF phase response (Fig.~\ref{fig:Interference}h) and the discrete dispersion steps (Fig.~\ref{fig:Interference}j) exhibit orthogonal mathematical signatures, we construct an overdetermined linear system from the measured $\tau_{\text{IQ}}(v)$ profile across a Doppler sweep. This enables the joint numerical reconstruction of both the instrument phase profile $\Phi(f)$ and the comb dispersion coefficient $\beta_2$, deterministically canceling both $\tau_{\text{RF}}$ and $\tau_{\text{dispersion}}$ to $<10$\,fs (see Methods for detailed derivations and $\tau_{\text{IQ}}$ curves). To ensure that residual calibration errors remain at the sub-femtosecond level, this numerical compensation is reinforced by dedicated physical optimizations of the detection hardware. Uncalibrated commercial detection chains typically exhibit group-delay ripples exceeding 10\,ns across the passband. Here, we engineered balanced photodetectors and custom low-pass filters specifically tailored for a maximally flat group-delay response, substantially flattening the raw RF phase profile and reducing the uncompensated nonlinear phase distortion by over an order of magnitude. Furthermore, because physical analog filters possess finite roll-off slopes, the interferograms are digitized at twice the comb repetition rate ($2f_{\text{r,A}} = 400$\,MHz) rather than the standard baseband rate, effectively suppressing out-of-band spectral aliasing.

With phase and dispersion errors systematically suppressed, the remaining frontier lies in the high-velocity dynamic regime ($V \ge 600$\,m/s). At these extreme speeds, both the I and Q channels suffer simultaneous envelope dilation (Fig.~\ref{fig:Interference}k), demonstrating that hardware interleaving alone cannot eliminate the Doppler-induced spectral breathing. To overcome this envelope distortion where hardware reaches its physical limit, we introduce an iterative interferogram reconstruction algorithm (detailed in Supplementary Information). Rather than relying on computationally prohibitive multidimensional CAF searches, the algorithm utilizes a coarse initial velocity estimate to reconstruct the static, zero-velocity interferogram via frequency-domain interpolation. The true arrival time is extracted from this restored frame, and a local constant-acceleration model updates the velocity across consecutive measurements. Theoretical simulations and benchmark tests confirm that two iterations suffice to suppress residual timing errors to the sub-femtosecond regime.

After applying these corrections to each one-way arrival time, the bidirectional clock offset is extracted by combining the counter-propagating links alongside the analytical non-reciprocal path correction ($Lv/2c^2 \approx 1$\,ns; see Supplementary Information). As summarized in Table~\ref{tab:error}, this integrated hardware-software framework suppresses all Doppler-induced systematic timing errors to within $\pm 10$\,fs across the entire 1.55\,km/s velocity range.

\begin{table}
    \centering
    \begin{tabular}{ccc}
        \hline
        \hline
        Errors & Bias if uncorrected & Bias after correction\\ 
        \hline
        Deformed interferogram (error of $T_{arrival}$) & $\sim$ps & $\lt0.1 fs$ (Theoretically)\\
        Nyquist dead zone & Signal blackout & Seamlessly eliminated\\
        Nonlinear radio frequency (RF) response ($\tau_{RF}$) & $\sim$10-1000 fs & $\lt10 fs$\\
        Comb chromatic dispersion ($\tau_{dispersion}$) & $\sim$10-100 fs & $\lt10 fs$\\
        Two-way non-reciprocal path ($\tau_{two-way}$) & $\sim$1\,ns & $\lt1 fs$\\
        \hline
        \hline
    \end{tabular}
    \caption{Estimated magnitudes of motion-induced timing errors and their post-compensation residuals in our experiment. All of them are compensated to within 10\,fs.}
    \label{tab:error}
\end{table}

\section{Experimental Demonstration}

\begin{figure}
    \centering
    \includegraphics[width=\linewidth]{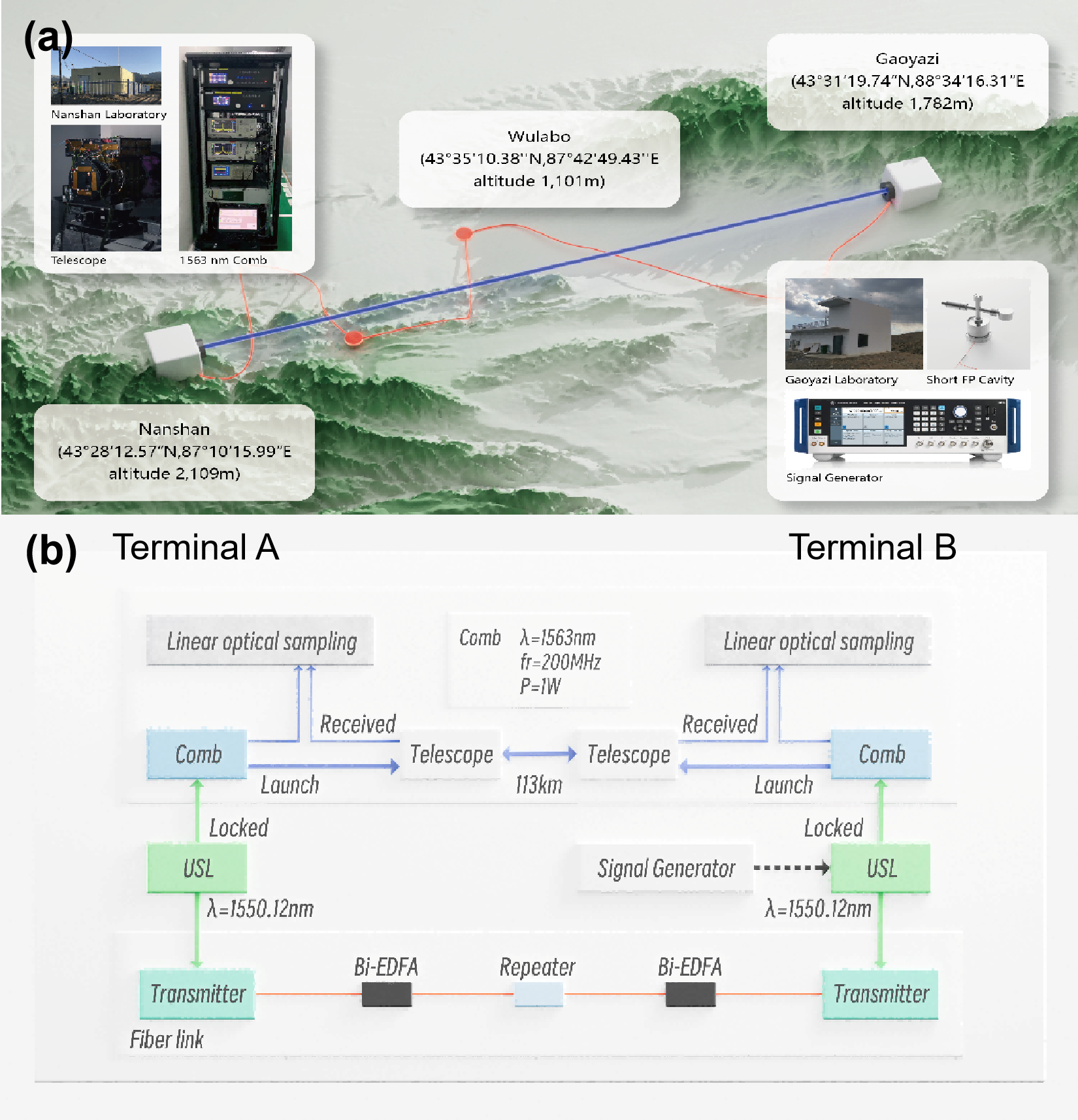}
    \caption{\textbf{Experimental setup for the 113-km Doppler-simulated time-frequency transfer. a,} Overview of the time-frequency transfer experiment The 113-km free-space optical link connects Nanshan (terminal A) and Gaoyazi (terminal B). Insets present field photographs of the 400-mm aperture optical telescopes, self-referenced frequency combs, short Fabry--P\'{e}rot (FP) cavities, and RF signal generators. 
    \textbf{b,} Detailed schematic of the dual-comb free-space link and the out-of-loop fiber verification link. At both ground terminals, an optical frequency comb (nominal repetition rate $f_{\text{r}} = 200$ MHz, optical power 1 W) is locked to an ultrastable optical reference ($\lambda = 1550.12$ nm). Satellite Doppler dynamics are emulated at Terminal B by modulating the output frequency of USL B via offset-frequency locking driven by a programmable signal source. Received optical comb signals are combined with local comb references in integrated linear optical sampling modules for precision arrival-time extraction. A parallel fiber-optic link equipped with bidirectional erbium-doped fiber amplifiers (Bi-EDFAs) and an optical repeater independently evaluates the clock offset between USL A and USL B.}
    \label{fig:setup}
\end{figure}

To validate the proposed compensation scheme and quantify the residual velocity-dependent timing errors, we conducted a 113-km free-space dual-comb transfer experiment with independent fiber-based out-of-loop verification in Urumqi, Xinjiang Province. The two terminals are located in Nanshan (terminal A) and Gaoyazi (terminal B), respectively as shown in Fig.~\ref{fig:setup}a. Fig.~\ref{fig:setup}b shows the basic setup: At each terminal, an optical frequency comb was stabilized in a self-referenced configuration to a 1550.12-nm ultrastable laser (USL). The comb signal was transmitted and received through a 400-mm aperture telescope, establishing a 113-km free-space channel. The specific configuration and atmospheric characteristics of the free-space channel are consistent with our previous work~\cite{shen2022free}, with a typical attenuation of 74\,dB. The optical comb had a wavelength range of 1558--1567\,nm and a power of 1\,W in this band. To avoid nonlinear effects induced by high peak power, the pulse width of the comb was broadened to approximately 80\,ps, with a dispersion of about 11\,ps/nm. The comb had a repetition rate of 200 MHz, and the difference in repetition rates between the two combs at terminals A and B was around 2\,kHz. The incoming comb signal interfered with the local comb in a linear optical sampling module. The resulting interference signal was detected, digitized, and processed to obtain one-way arrival-time measurements. The dual-comb comparison ultimately yielded the phase difference between USL A and USL B. 

To evaluate the performance of the dual-comb comparison link, we used a fiber-optic frequency transfer link as an out-of-loop verification. This fiber-optic frequency transfer link included two bidirectional erbium-doped fiber amplifiers (Bi-EDFAs) and a repeater. The repeater located in Wulabo divided the fiber link into two segments of 102\,km and 107\,km each, improving the noise suppression capability and making it better suited for our almost entirely aerial fiber link. At Terminal B, the fiber-delivered USL A signal was beat with USL B, and after a 16-fold frequency division, the phase was recorded by a $\Pi$-type frequency counter. The reference of the frequency counter is synchronized with the repetition rate of the comb at Terminal B.

Relative platform motion was optically emulated at Terminal B by locking USL B in an offset-frequency Pound--Drever--Hall (PDH) configuration on its $-1$ sideband, rendering its output frequency continuously tunable according to $f_{\text{CW,B}} = f_{\text{cavity}} + f_{\text{EOM}}$. By dynamically modulating $f_{\text{EOM}}$ via a programmable signal source across a 1 GHz span (from 500 MHz to 1.5 GHz, bounded by photodiode bandwidths and frequency-divider limits), we generated relative optical carrier translations equivalent to velocities of up to 1.55 km/s. It was noted that as $f_{\text{EOM}}$ varies, the local sampling clock and frequency counter timebase at Terminal B experience proportional frequency drifts. Left uncompensated, these timebase shifts misalign the two-way timestamps required for bidirectional link-noise cancellation. We therefore implemented an additional post-processing timestamp correction step at Terminal B to preserve data-pairing alignment and maintain the cancellation of common-mode link noise (see Supplemental Material for details).

\begin{figure}
    \centering
    \includegraphics[width=\linewidth]{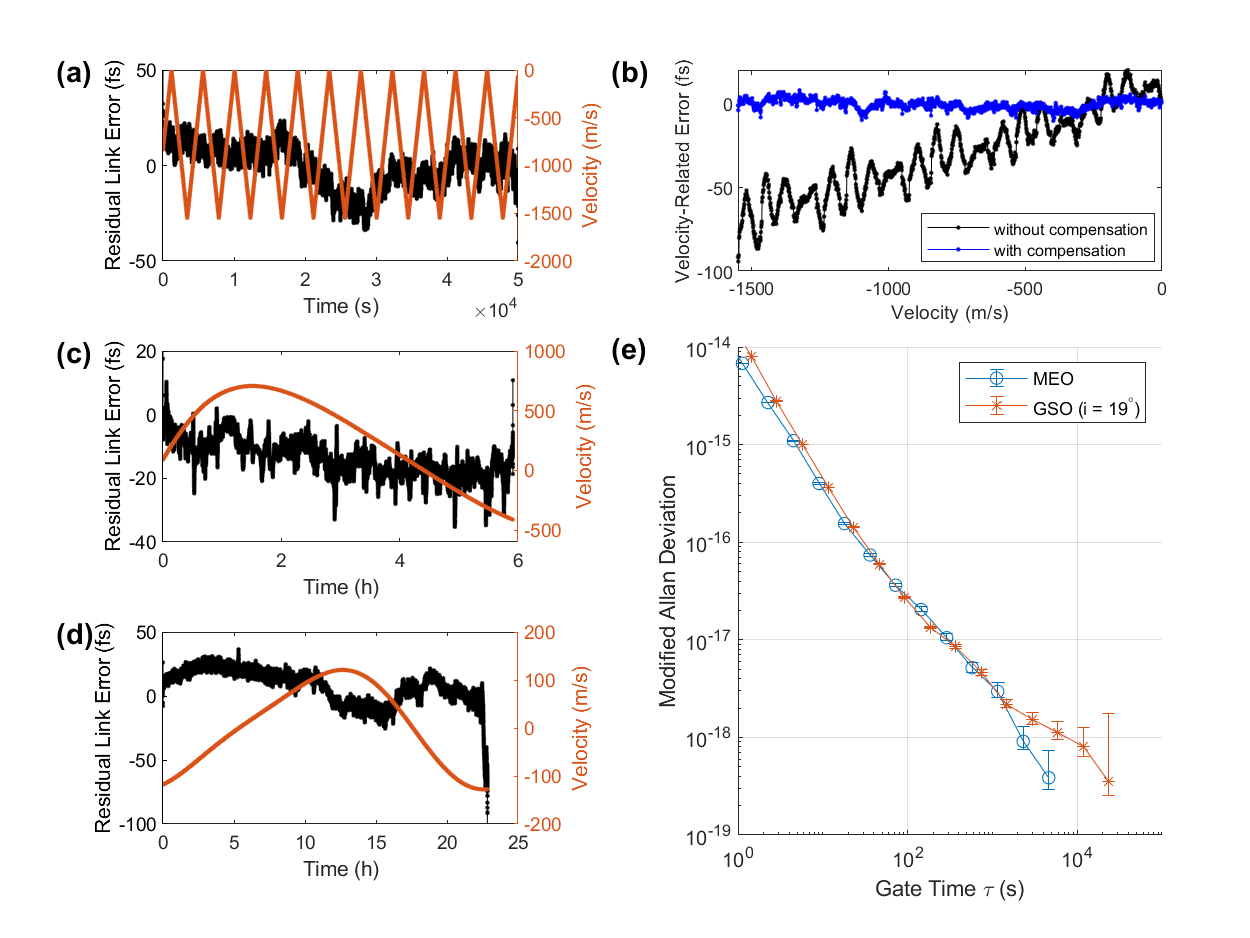}
    \caption{
    \textbf{Performance evaluation of Doppler-compensated time-frequency transfer. a,}  Doppler velocity sweep up to 1.55 km/s and corresponding clock offset. The orange curve displays the triangular velocity profile swept continuously from 0 to $-1.55$ km/s (right axis), while the black curve shows the smoothed clock offset after background drift subtraction (left axis).
    \textbf{b,} Velocity-dependent timing errors evaluated across the entire 1.55 km/s speed span. Black curve: uncompensated systematic error, exhibiting biases exceeding 100 fs. Blue curve: residual error after applying compensation, suppressed to within $\pm 10$ fs. 
    \textbf{c,} Simulated MEO satellite trajectory over a 6-h visibility window. Orange curve shows the relative radial velocity variation ($-400$ m/s to $+700$ m/s); black curve represents the smoothed clock offset.
    \textbf{d,} Simulated GSO trajectory (inclination $i = 19^\circ$) over a 23-h window, featuring velocity variations within $\pm 150\text{ m/s}$ (orange) and the corresponding clock offset (black).
    \textbf{e,} Fractional frequency instability (modified Allan deviation) for the simulated MEO (blue circles) and GSO (orange triangles) links, both reaching $10^{-19}$ at averaging time. Error bars represent $1\sigma$ confidence intervals.
    }
    \label{fig:r2}
\end{figure}

We first characterized and quantified the motion-dependent systematic timing errors across the entire velocity range. Defining receding relative motion as positive (where the received optical frequency decreases), $f_{\text{EOM}}$ was modulated in a triangular sweep profile from 500\,MHz to 1.5\,GHz at a rate of 450\,kHz/s with a 2-ms step interval. Defining the initial state ($f_{\text{EOM}} = 500\text{ MHz}$) as zero velocity, the increasing output frequency $f_{\text{CW,B}}$ translated to an equivalent velocity range from 0 to $-1.55\text{ km/s}$ from Terminal A's perspective, and from 0 to $+1.55\text{ km/s}$ from Terminal B's perspective. Adopting Terminal A as the system reference, Fig.~\ref{fig:r2}a displays the dynamic triangular velocity profile alongside the continuous out-of-loop comparison offset, obtained by subtracting the independent fiber verification measurements from the free-space dual-comb data. Because ambient environmental fluctuations across the open-air channel induce a slow background thermal drift, we modeled this baseline variation via piecewise cubic interpolation between zero-velocity baseline points recorded at the start and end of each sweep cycle (Fig.~\ref{fig:r2}a, black curve). After subtracting this background drift, statistical analysis binned at 1\,m/s velocity intervals revealed that the uncompensated velocity-dependent systematic error exceeded 100\,fs (Fig.~\ref{fig:r2}b, black curve). Applying our joint dual-channel calibration and iterative reconstruction framework suppressed the residual systematic timing error to within $\pm 10$\,fs across the entire 1.55\,km/s velocity span (Fig.~\ref{fig:r2}b, blue curve)---representing a tenfold reduction in systematic bias.

Finally, we programmed $f_{\text{EOM}}$ to execute velocity profiles corresponding to typical GNSS operational constellations: an MEO satellite (radial velocity range $-400\text{ m/s}$ to $+700\text{ m/s}$) and an inclined GSO satellite ($i = 19^\circ$, radial velocity range within $\pm 150\text{ m/s}$). Setting the central $f_{\text{EOM}}$ frequency to 1.0\,GHz provided symmetric margins to accommodate both receding and approaching orbital dynamics. Crucially, experimentally validating this dynamic inclined GSO trajectory inherently confirms operability for the entire GSO orbital regime, trivially encompassing stationary GEO links ($v = 0\text{ m/s}$). Over continuous measurements, the emulated links achieved fractional frequency instabilities (modified Allan deviations) of $3.9 \times 10^{-19}$ at 4,600\,s for MEO and $3.6 \times 10^{-19}$ at 23,600\,s for GSO (Fig.~\ref{fig:r2}e).

\section{Discussion and Outlook}

In summary, we have demonstrated high-precision free-space time-frequency transfer over a 113-km link, successfully overcoming the physical barriers imposed by high-velocity platform motion. By combining a quadrant-shifted dual-channel sampling architecture with an iterative interferogram reconstruction algorithm, residual systematic timing errors are suppressed to within $\pm 10$\,fs across the entire 1.55\,km/s velocity excursion. The fractional frequency instabilities achieved at the $10^{-19}$ level for simulated MEO and GSO trajectories confirm that this system satisfies the rigorous requirements for intercontinental comparisons of state-of-the-art optical clocks.

The methodology demonstrated here is scalable to higher-velocity scenarios, such as Low Earth Orbit (LEO) or Earth-Moon links. In such regimes, larger Doppler shifts can drive the received repetition rate to approach that of the local oscillator, causing linear optical sampling to encounter repetition-rate degeneracy and spectral folding. This challenge can be fundamentally bypassed by allocating a sufficiently wide initial repetition-rate offset ($\Delta f_{\text{r}}$). Ensuring that the maximum Doppler shift does not invert or equate the repetition rates preserves continuous sampling, which, combined with our iterative reconstruction algorithm, enables robust operation over hyper-dynamic time-frequency transfer links.

By resolving these motion-induced impairments, this work transitions optical time-frequency transfer from quasi-static demonstrations to dynamic orbital platforms, unlocking global-scale applications. It removes the major technical obstacle to a satellite-based femtosecond-precision network and paves the way for the future redefinition of the SI second, ultra-precise relativistic geodesy, and high-sensitivity tests of fundamental physics in space.

\vspace{1em}
\noindent\textbf{Note added:} During the preparation of this manuscript, we noted a concurrent benchtop study demonstrating Doppler emulation up to 620\,m/s in a controlled laboratory environment without dynamic error compensation or turbulent free-space propagation~\cite{bigelow2026development}.

\section*{Methods}

\subsection{Doppler simulation configuration}

When a signal is affected by the Doppler effect, each frequency component experiences a frequency shift proportional to the relative velocity. For a single-frequency continuous-wave (CW) reference laser, the Doppler shift is simply given by \(\Delta f_{\text{CW}} = f_{\text{CW}} \frac{v}{c}\), where \(v\) is the relative velocity and \(c\) is the speed of light. To faithfully emulate this velocity-dependent frequency translation on the ground without physical motion, we employ a broadly tunable ultrastable laser (USL) stabilized via an offset-frequency Pound-Drever-Hall (PDH) locking scheme.

\begin{figure}
    \centering
    \includegraphics[width=\linewidth]{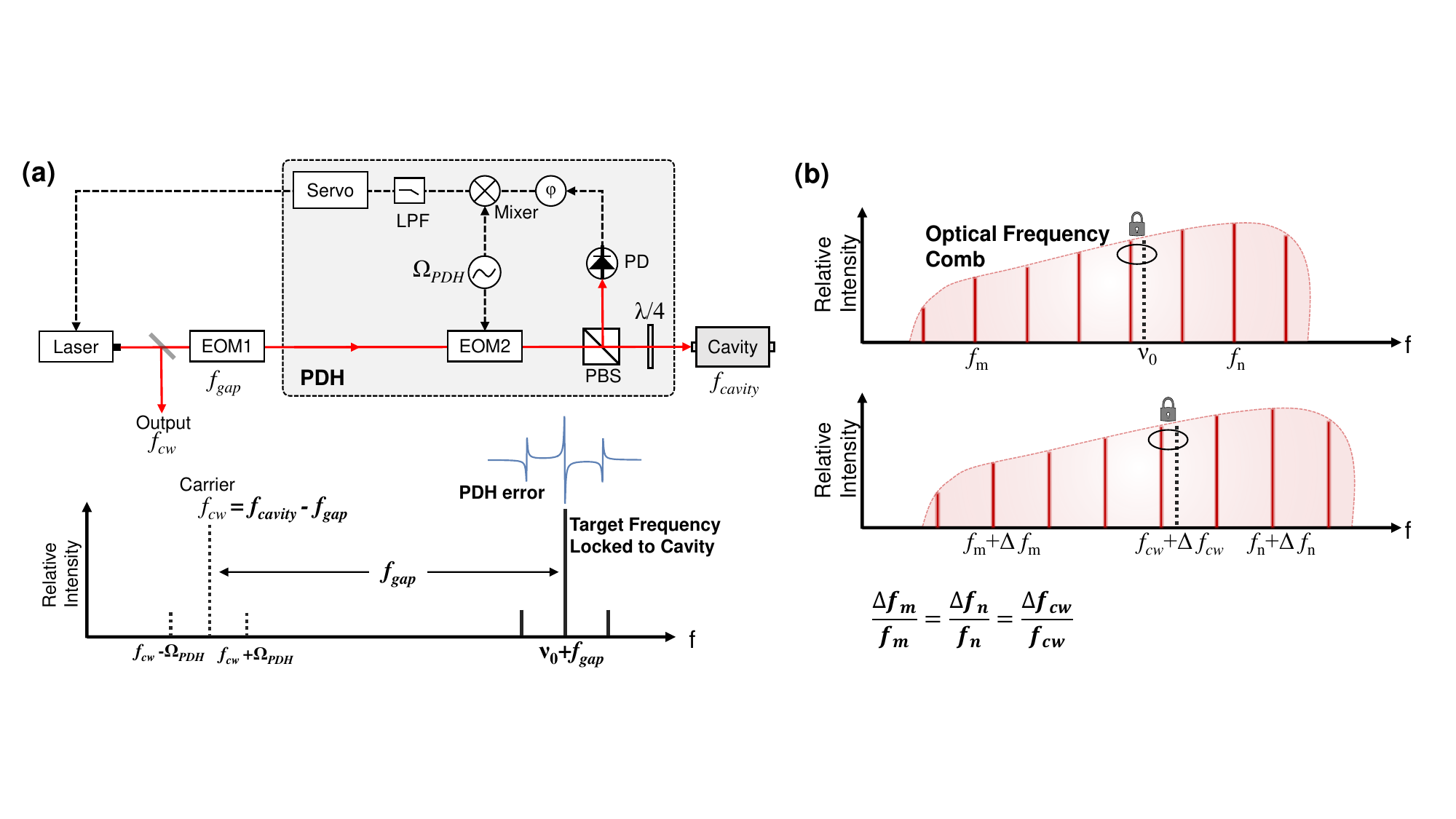}
    \caption{
    \textbf{Ground-based Doppler simulation scheme and offset-frequency locking architecture. a,} Schematic of the offset-frequency Pound--Drever--Hall (PDH) locking setup. A wideband radio-frequency tuning signal ($f_{\text{EOM}}$) is applied via electro-optic modulation (EOM1) to generate sidebands on the seed laser output. One sideband is stabilized to a high-finesse optical cavity using standard PDH modulation ($\Omega_{\text{PDH}}$), locking the output CW carrier frequency to $f_{\text{CW}} = f_{\text{cavity}} \pm f_{\text{EOM}}$. Dynamically sweeping $f_{\text{EOM}}$ enables continuous, wide-range optical frequency tuning without physical movement. PBS, polarizing beam splitter; PD, photodiode; LPF, low-pass filter; Servo, feedback loop controller.
    \textbf{b,} Principle of Doppler-equivalent optical frequency comb spectrum scaling. Under a self-referenced locking configuration, tuning the continuous-wave (CW) reference laser frequency ($f_{\text{CW}}$) induces a strictly proportional frequency shift across all comb teeth, satisfying $\Delta f_{\text{m}}/f_{\text{m}} = \Delta f_{\text{n}}/f_{\text{n}} = \Delta f_{\text{CW}}/f_{\text{CW}}$. This synchronized "breathing" mode faithfully reproduces the spectral stretching behavior caused by physical relativistic Doppler motion.}
    \label{fig:EOM}
\end{figure}

Fig.~\ref{fig:EOM}a shows the principle of frequency shift. In a conventional PDH locking scheme, the laser is phase-modulated by an electro-optic modulator (EOM) to generate sidebands with a frequency $\Omega_{\text{PDH}}$, and the reflected signal from the cavity is demodulated with the same frequency to produce an error signal for frequency stabilization. Ultimately, the output laser frequency \(f_{\text{CW}}\) is made to match the cavity transmission frequency \(f_{\text{cavity}}\). To enable wide-range frequency tuning, we introduce an additional frequency-shift signal \(f_{\text{EOM}}\) into a second EOM. This generates the \(\pm 1\text{st}\)-order sidebands. \(f_{\text{EOM}}\) is much greater than $\Omega_{\text{PDH}}$ to ensure that no sideband overlaps with the carrier. Then, one sideband is locked to the cavity using the conventional PDH method, and the carrier is output to the comb. The carrier frequency is given by \(f_{\text{CW}} = f_{\text{cavity}} \pm f_{\text{EOM}}\). The plus and minus signs correspond to different sideband lockings. In our experiment, to ensure that the CW frequency difference between Terminal A and Terminal B always stays within the 300\,MHz to 2\,GHz range imposed by the electronics system bandwidth over the entire frequency-shifting range, we lock the $-1$ sideband, so that the output CW frequency at Terminal B is given by \(f_{\text{CW,B}} = f_{\text{cavity}} + f_{\text{EOM}}\).

For an optical frequency comb, however, the Doppler effect manifests as a proportional scaling of the entire optical spectrum rather than a uniform translation. To correctly emulate this behavior, we adopt a self-referenced locking scheme. The optical comb is stabilized by independently locking the CEO frequency ($f_{\text{CEO}}$) and the beat frequency ($f_{\text{beat}}$) between one comb tooth $m$ and the reference CW. Both $f_{\text{CEO}}$ and $f_{\text{beat}}$ are synchronized with the comb repetition rate ($f_{\text{rep}}$). This implies $f_{\text{CEO}} = r_{\text{CEO}} f_{\text{rep}}$ and $f_{\text{beat}} = r_{\text{beat}} f_{\text{rep}}$, where $r_{\text{CEO}}$ and $r_{\text{beat}}$ are fractional coefficients. The repetition frequency of the comb is then uniquely determined by the CW laser frequency as follows:

\begin{equation}
    f_{\text{CW}}=f_{\text{CEO}}+m f_{\text{rep}}+f_{\text{beat}}=(m+r_{\text{CEO}}+r_{\text{beat}})f_{\text{rep}},
\end{equation}
\begin{equation}
    f_{\text{rep}}=\frac{f_{\text{CW}}}{m+r_{\text{CEO}}+r_{\text{beat}}}.
\end{equation}

Thus, the frequency of the $n$-th comb tooth can be written as
\begin{equation}
f_\text{n}=(n+r_{\text{CEO}})f_{\text{rep}}=f_{\text{CW}}\frac{n+r_{\text{CEO}}}{m+r_{\text{CEO}}+r_{\text{beat}}}.
\end{equation}
When \(f_{\text{CW}}\) changes by \(\Delta f_{\text{CW}}\), the corresponding shift of the \(n\)-th comb tooth is \(\Delta f_n = f_n \frac{\Delta f_{\text{CW}}}{f_{\text{CW}}}\), meaning every comb tooth is strictly proportional to the CW laser frequency and scales synchronously. By tuning \(f_{\text{EOM}}\) and thereby shifting \(f_{\text{CW,B}}\), the full frequency comb at Terminal B is driven into a synchronized breathing mode, perfectly emulating the Doppler stretching behavior over the entire target velocity range. The equivalent speed can be given by $v_{\text{sim}} = c \frac{\Delta f}{f_{\text{CW}}}$, as shown in Fig.~\ref{fig:EOM}b.

Compared to the actual Doppler effect, this high-speed Doppler simulation method exhibits identical behavior only for the signal path from Site B to Site A. For the reverse path (Site A to Site B), the frequency of the signal comb transmitted from A remains unchanged. Instead, it is the frequency of the local comb at Site B, involved in the linear optical sampling process, that is altered. From the perspective of Site B, this is equivalent to the comb signal arriving from Site A experiencing a frequency shift in the opposite direction. Consequently, the equivalent velocity for this path is opposite to that of the B-to-A direction:
\begin{equation}
\begin{aligned}
    v_{\text{sim,A}} = &c \frac{\Delta f}{f_{\text{CW,B}}},\\
    v_{\text{sim,B}} = &c \frac{f_{\text{CW,A}}}{f_{\text{CW,B}}-\Delta f} \frac{\Delta f}{f_{\text{CW,A}}},\\
    v_{\text{sim,B}} = &c \frac{\Delta f}{f_{\text{CW,B}}-\Delta f}=c \frac{\Delta f/f_{\text{CW,B}}}{1-\Delta f/f_{\text{CW,B}}}=c \frac{v_{\text{sim,A}}/c}{1-v_{\text{sim,A}}/c}\approx v_{\text{sim,A}}, (\text{for }v_\text{sim} \ll c).
\end{aligned}
\end{equation}
The factor $\frac{f_{\text{CW,A}}}{f_{\text{CW,B}}-\Delta f}$ in $v_{\text{sim,B}}$
stems from the choice of frequency reference: from Terminal B's viewpoint,
that reference ought to be the current frequency of the CW laser at Terminal B.

This leads to a minor discrepancy between the formula for the calculation of the two-way clock offset in the actual Doppler scenario, specifically manifested in the sign of the bidirectional asymmetry term $\frac{Lv}{2c^2}$. The minor discrepancy does not compromise the validity of our high-speed Doppler simulation.

\subsection{Detailed explanation of Motion-Induced One-Way Errors}

In the main text, we qualitatively outlined the principal Doppler-induced timing errors in dual-comb linear optical sampling. Here, we provide a rigorous model that underpins those effects. The electric fields of the local comb (A) and the signal comb (B) are
\begin{equation}
\begin{aligned}
E_{\text{A}}(t) &= e^{i2\pi \nu_{\text{A}} t} \sum_m E_{\text{A},m} e^{i(m2\pi f_{\text{r,A}}t + \phi_{\text{A},m})},\\
E_{\text{B}}(t) &= e^{i2\pi \nu_{\text{B}} (t-T_{\text{link}})} \sum_n E_{\text{B},n} e^{i[n2\pi f_{\text{r,B}}(t-T_{\text{link}}) + \phi_{\text{B},n}]},
\end{aligned}
\end{equation}
where $\nu_{\text{A}}$ and $\nu_{\text{B}}$ denote the optical carrier frequencies of the central teeth for combs A and B, respectively, with $\nu_{\text{B}}$ being the line most adjacent to $\nu_{\text{A}}$. The phase terms $\phi_{\text{A},m}$ and $\phi_{\text{B},n}$ represent the intrinsic chromatic dispersion of the respective combs. Owing to the slight offset between their repetition rates ($\Delta f_{\text{r}} = f_{\text{r,B}} - f_{\text{r,A}}$), the optical pulses from comb B periodically scan through those of comb A. This cross-correlation cycle yields a periodic time-domain multi-heterodyne signal designated as the interferogram: 
\begin{equation}
I_{\text{A}}(t) = R(t) * \text{Im}\left[E_{\text{A}}^*(t) E_{\text{B}}(t)\right],
\end{equation}
where $R(t)$ represents the impulse response function of the detection electronics, and $*$ denotes the convolution operator.  Multi-heterodyne mixing between all mutually coherent comb lines generates a dense radio-frequency (RF) spectrum, where each discrete RF line encodes distinct timing information. Conventionally, a low-pass filter (LPF) with a sharp cutoff at $f_{\text{r,A}}/2$ isolates the primary interference signal produced by the closest comb-tooth pairings. In the time domain, this band-limited interferogram can be explicitly expressed as: 
\begin{equation}
\begin{aligned}
I_{\text{A}}(t) =\text{Im}\{& \exp\bigl[ i2\pi (\nu_{\text{B}}-\nu_{\text{A}})t - i2\pi \nu_{\text{B}} T_{\text{link}} \bigr] \\
\times &\sum_m \exp\bigl[ i2\pi m\Delta f_{\text{r}} t - i2\pi m f_{\text{r,B}} T_{\text{link}} + i(\phi_{\text{B},m} - \phi_{\text{A},m}) + i\angle R(\nu_{\text{B}}-\nu_{\text{A}}+m\Delta f_{\text{r}}) \bigr]\},
\end{aligned}
\end{equation}
where \(\angle R\) is the frequency-dependent phase response of the electronic chain. To solve for the timing information, we first transform the interferograms to the frequency domain, and the angle of each frequency $f$ is denoted as $\angle I(f,n)$ for the $n$-th interferogram. Defining the optical carrier offset as $\Delta \nu_{\text{AB}} = \nu_{\text{B}} - \nu_{\text{A}}$ and the link magnification factor as $M = f_{\text{r,B}} / \Delta f_{\text{r}}$, we have:
\begin{equation}\label{eq:phase}
\angle I(f,n) = \angle I(f,0) - 2\pi \nu_{\text{B}} \Delta T_{\text{link}} + 2\pi M(f - \Delta \nu_{\text{AB}}) \Delta T_{\text{link}}.
\end{equation}
Here $\angle I(f,0)$ means the phase of the initial interferogram and is used as a reference. The phase difference of $\angle I(f,n)$ and $\angle I(f,0)$ has a linear relationship with frequency $f$. The correlation slope is $s=2\pi M \Delta T_{\text{link}}$. So the change in one-way arrival times since the initial interferogram, $\Delta T_{\text{link}}$, is derived from the slope of the phase difference with frequency $f$. Crucially, both the optical chromatic dispersion ($\phi_{\text{B},m} - \phi_{\text{A},m}$) and the electronic RF dispersion $\angle R$ are perfectly common-mode canceled. This elimination occurs because the spectral comparison between the $n$-th interferogram $I(f,n)$ and the initial reference $I(f,0)$ isolates identical comb-tooth pairings at the same down-converted frequencies. However, this no longer holds under the Doppler effect.

Under a Doppler shift with a relative radial velocity \(v\), the following error mechanisms emerge:

\textbf{Nonlinear RF response.} At low velocities of several tens of m/s, the RF spectrum of the interferogram undergoes a rigid frequency translation by \(f_{\text{shift}} \approx \nu_{\text{B}} \frac{v}{c}\). While the cross-ambiguity function (CAF) algorithm~\cite{sinclair2019femtosecond} can realign the spectra, the frequency offset breaks the common-mode cancellation of the electronic phase response \(\angle R\). The residual term \(\angle R(f,n) - \angle R(f,0)\) couples the RF group-delay dispersion into a velocity-dependent timing error that can reach hundreds of femtoseconds in our system. Although this error dominates at low velocities, it persists across the entire velocity range. Typically, the nonlinearity of RF electronics is characterized by the group delay \(T_\text{g}(f)\), which physically represents the additional delay experienced by a narrow-band signal centered at frequency \(f\). The resulting timing error in dual-comb measurements, \(\tau_{RF}\), can be expressed as \(\tau_\text{RF} \approx T_\text{g}(f) / M\), where the approximation accounts for the finite spectral bandwidth of the interferogram. In our system, the uncalibrated RF group delay across the photodetector and hardware low-pass filters varies by more than 10\,ns, precipitating a raw timing error on the order of hundreds of femtoseconds.

\textbf{Nyquist dead zone.} When the center frequency of the multi-heterodyne spectrum approaches 0 or \(f_{\text{r}}/2\), the positive and negative frequency components fold onto each other, rendering the phase profile in Eq.~\eqref{eq:phase} ill-defined and making phase-slope extraction impossible. These dead zones recur periodically every $f_{\text{r}}/2\approx 100$\,MHz of carrier-frequency translation, corresponding to velocity intervals of \(\sim 155\)\,m/s in our experiment.

\textbf{Comb chromatic dispersion.} Beyond the dead zone of the 155\,m/s threshold, an effective interferogram re-emerges. Now the frequency shift \(f_{\text{shift}} \approx \nu_{\text{B}} \frac{v}{c}\) exceeds \(f_{\text{r,A}}/2\), so the original interference signal of the correctly paired comb tooth lies outside the low-pass filter passband. The signal passed by the filter therefore originates from a mismatched pair of optical comb lines. Specifically, the dynamic phase profile \(\angle I(f,n)\) accumulates a mismatched differential optical phase (e.g., \(\phi_{\text{B,m}} - \phi_{\text{A,m-1}}\)) compared to the matched pairing (\(\phi_{\text{B},m} - \phi_{\text{A,m}}\)) preserved in the reference \(\angle I(f,0)\). Here we present a detailed derivation of \(\tau_{\text{dispersion}}\). Let the tooth-index offset be \(k\). We then have

\begin{equation}
\angle I(f,n) = \angle I(f,0) - 2\pi \nu_{\text{B}} T_{\text{link}} + 2\pi M(f - \Delta \nu_{\text{AB}}) T_{\text{link}} + \phi_{\text{A},m} - \phi_{\text{A},m - k}.
\end{equation}

The extra phase term \(\phi_{\text{A},m} - \phi_{\text{A},m - k}\) can be expressed through the comb dispersion coefficient \(\beta_2\) as
\begin{equation}
\begin{aligned}
\phi_{\text{A},m} - \phi_{\text{A},m - k} = &\frac{1}{2}\beta_2(2\pi m f_\text{r,A})^2-\frac{1}{2}\beta_2[2\pi (m-k) f_\text{r,A}]^2\\
\approx&4\pi^2m k \beta_2f_\text{r,A}^2.
\end{aligned}
\end{equation}

Noting that \(m f_{\text{r,A}} = m (M-1)\Delta f_{\text{r}} = (M-1)(f - \Delta \nu_{\text{AB}}) \approx M(f - \Delta \nu_{\text{AB}})\), we obtain
\begin{equation}
\begin{aligned}
\angle I(f,n) -\angle I(f,0) = &- 2\pi \nu_{\text{B}} T_{\text{link}} + 2\pi M(f - \Delta \nu_{\text{AB}}) T_{\text{link}} + 4\pi^2k \beta_2 M(f-\Delta \nu_{\text{AB}}) f_\text{r,A}\\
\approx&- 2\pi \nu_{\text{B}} T_{\text{link}} + 2\pi M(f - \Delta \nu_{\text{AB}})(T_{\text{link}}+2\pi k \beta_2 f_\text{r,A}).
\end{aligned}
\end{equation}
Hence, the effect of comb chromatic dispersion on the timing measurement is \(\tau_{\text{dispersion}} = 2\pi k \beta_2 f_{\text{r,A}}\). As the velocity increases, the tooth misalignment index \(k\) grows, and so does this error. In our system, the comb dispersion is 11~ps/nm, corresponding to a timing error of 18~fs per misaligned tooth. Over the full 1.55~km/s range, up to 5 teeth can be misaligned, leading to a worst-case error approaching 100~fs.

\textbf{Deformed interferogram.} At velocities exceeding \(\sim 600\)\,m/s, the relativistic Doppler effect significantly alters the repetition-rate offset:
\begin{equation}
\Delta f_{\text{r}}' = f_{\text{r,sig}}\frac{c-v}{c} - f_{\text{r,local}}.
\end{equation}
For a nominal \(\Delta f_{\text{r}} = 2\)\,kHz, this causes a \(\sim 20\%\) variation in the RF spectral width, compressing or dilating the interferogram envelope. Under this distortion, the CAF assumption of a constant \(\Delta f_{\text{r}}\) breaks down, and the extraction of \(T_{\text{link}}\) via Eq.~\eqref{eq:phase} yields timing deviations of hundreds of femtoseconds or more.

%\textbf{Two-way non-reciprocal path.} In a two-way time--frequency transfer, the signal propagates from A to B and from B to A. Because the satellite (or simulated velocity) is moving during the round-trip, the up-link and down-link path lengths are not equal. This asymmetry introduces a non-reciprocal timing bias given by $\tau_{two-way} = L v / (2 c^2)$, where $L$ is the instantaneous link distance and $v$ is the relative radial velocity. Although this term is straightforward to calculate, it is several nanoseconds in our 113-km link at $v = 1.55$~km/s and must be subtracted from the two-way difference to obtain the true clock offset.

\subsection{Joint correction of RF response and comb dispersion by dual-channel sampling}

The dual-channel architecture enables quantitative compensation for the errors introduced by the nonlinear RF response and the comb chromatic dispersion. Theoretically, after all errors are properly compensated, the one-way arrival times derived from the two channels should be consistent with each other, differing only by the artificial non-symmetric delay introduced in the optical paths:
\begin{equation}
    \begin{aligned}
    \tau_I^{\text{comp}}(v)&=\tau_Q^{\text{comp}}(v)-\tau_{\text{delay}},\\
    \tau_I^{\text{comp}}(v)&=\tau_I(v) - \tau_{\text{RF}}^I(v) - \tau_{\text{disp}}^I(v),\\
    \tau_Q^{\text{comp}}(v)&=\tau_Q(v) - \tau_{\text{RF}}^Q(v) - \tau_{\text{disp}}^Q(v),\\
    \tau_Q(v) - \tau_I(v) &=\tau_{\text{delay}}+\tau_{\text{RF}}^Q(v) + \tau_{\text{disp}}^Q(v)-\tau_{\text{RF}}^I(v) - \tau_{\text{disp}}^I(v)
    \end{aligned}
\end{equation}
where $\tau_{\text{RF}}^{\text{I,Q}}(v)$ and $\tau_{\text{disp}}^{\text{I,Q}}(v)$ are the RF response-induced timing error and optical comb dispersion-induced timing error for Channels~I and Q, respectively. In practice, the raw differential arrival time between the two channels, $\tau_{\text{Q}}(v) - \tau_{\text{I}}(v)$, is dominated by the large static delay line ($\tau_{\text{delay}} \approx 1.5$\,ns), superimposed with subtle, femtosecond-level velocity-dependent variations. To isolate these velocity-dependent dynamics, we subtract the static DC baseline and define the zero-centered differential timing metric as:
\begin{equation}
    \tau_{\text{IQ}}(v) = \tau_{\text{Q}}(v) - \tau_{\text{I}}(v) - \tau_{\text{delay}}',
\end{equation}
where the nominal baseline $\tau_{\text{delay}}' = \langle \tau_{\text{Q}}(v) - \tau_{\text{I}}(v) \rangle$ is obtained by averaging the raw inter-channel differential delay over a symmetric Doppler sweep. This centering operation eliminates the large constant offset, ensuring that $\tau_{\text{IQ}}(v)$ directly captures the anti-symmetric RF phase nonlinearity and discrete comb dispersion steps, while any minute residual baseline offset ($\tau_{\text{delay}} - \tau_{\text{delay}}'$) is deterministically resolved in the subsequent joint inversion.

The characteristics of the \(\tau_{IQ}\) curve provide clear evidence for the combined effects of the nonlinear RF response and chromatic dispersion. Under the assumption of constant repetition rates, the spectral width of the interferogram remains unchanged, with only its center position shifting as a function of velocity. The center frequencies of the spectrum for Channel I and Channel Q are denoted as \(f_I\) and \(f_Q\), respectively. Positive values of \(f_I\) and \(f_Q\) are defined as those for which the optical comb-tooth frequency from Terminal B is lower than that from Terminal A. For Terminal A, the frequency relation is given by \(f_Q = f_I + f_{\text{AOM}}\). It is clearer to show the characteristics of the \(\tau_{IQ}\) curve by changing the independent variable from speed \(v\) to the center frequency of channel I \(f_I\). And we have 
\begin{equation}
    \tau_{\text{IQ}}(f_{\text{I}}) = \tau_{\text{Q}}(f_{\text{Q}}) - \tau_{\text{I}}(f_{\text{I}}) - \tau_{\text{delay}}' = \tau_{\text{delay}} - \tau_{\text{delay}}' + \tau_{\text{RF}}(f_{\text{Q}}) + \tau_{\text{disp}}(f_{\text{Q}}) - \tau_{\text{RF}}(f_{\text{I}}) - \tau_{\text{disp}}(f_{\text{I}})
\end{equation}

\begin{figure}
    \centering
    \includegraphics[width=\linewidth]{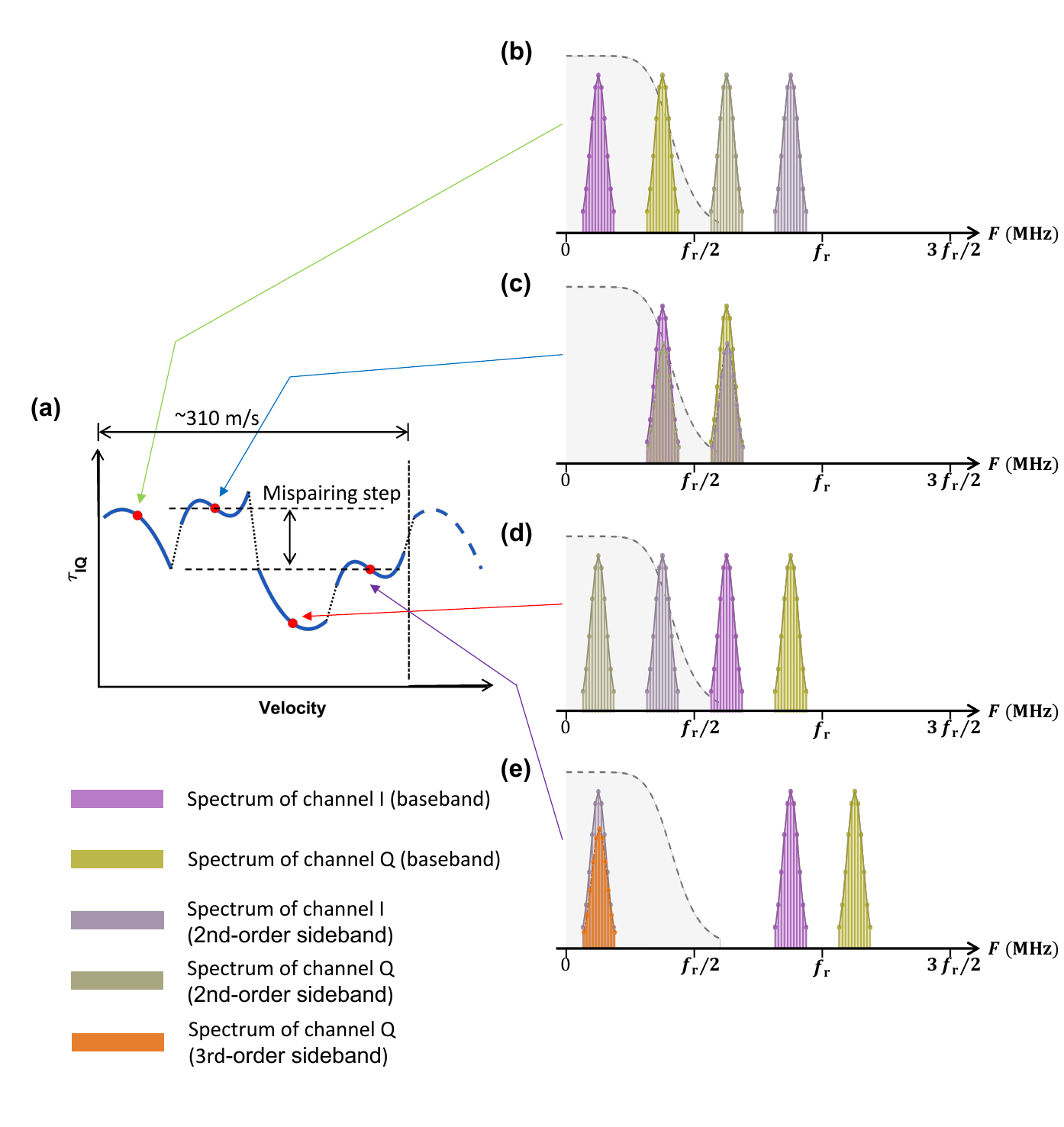}
    \caption{\textbf{Relative inter-channel time delay difference \(\tau_{\text{IQ}}\) versus Doppler velocity and the underlying multi-heterodyne spectral dynamics.} 
    \textbf{a,} Segmented \(\tau_{\text{IQ}}\) profile as a function of Doppler velocity across one repetition-rate period (\(\sim 310\)~m/s, with the dashed blue curve indicating continuation into the second period). The intra-segment curvature and central anti-symmetry originate from the nonlinear phase response of the detection system combined with Nyquist folding, while the discrete step-like discontinuity (annotated as ``Mispairing step'' between the centrosymmetric centers of the second and fourth segments) stems from optical comb-tooth mispairing. 
    \textbf{b--e,} Multi-heterodyne radio-frequency (RF) spectra of Channel~I and Channel~Q evaluated at four representative operating velocities (linked by colored arrows to the red markers in \textbf{a}).}
    \label{fig:IQ_spectrum}
\end{figure}

Fig.~\ref{fig:IQ_spectrum} illustrates the underlying mechanism governing the relative time delay difference \(\tau_{IQ}\) across the frequency \(f_I\).

When \(f_I < f_{\text{rep}}/4\), the \(\tau_{IQ}\) curve exhibits central anti-symmetry about the frequency point \(f_I = f_{\text{AOM}}/2\). This symmetry center corresponds to the condition where the spectra of Channel I and Channel Q are centered symmetrically at \(-f_{\text{AOM}}/2\) and \(f_{\text{AOM}}/2\), respectively. The explicit derivation of this anti-symmetry is expressed as follows. Setting \(f_{I1} = -(f_{I2} + f_{\text{AOM}})\), we have \(-(f_{I1} + f_{\text{AOM}}) = f_{I2}\). Since both \(f_I\) and \(f_Q\) satisfy \(|f|<f_{\text{rep}}/2\), their comb tooth pairings are the same, leading \(\tau_{\text{disp}}(f_I)=\tau_{\text{disp}}(f_Q)\). So we have the relation:

\begin{equation}
\begin{aligned}
    \tau_{IQ}(f_{I2})=&\tau_{\text{delay}}- \tau_{\text{delay}}'+\tau_{\text{RF}}(f_{I2}+f_{\text{AOM}}) -\tau_{\text{RF}}(f_{I2})\\
    =&\tau_{\text{delay}}- \tau_{\text{delay}}'+\tau_{\text{RF}}(-f_{I1})-\tau_{\text{RF}}(-(f_{I1}+f_{\text{AOM}}))\\
    =&\tau_{\text{delay}}- \tau_{\text{delay}}'+\tau_{\text{RF}}(f_{I1})-\tau_{\text{RF}}(f_{I1}+f_{\text{AOM}})\\
    =&\tau_{\text{delay}}- \tau_{\text{delay}}'-\tau_{IQ}(f_{I1}), \quad f_{I1}<f_{\text{rep}}/4.
\end{aligned}
\end{equation}

When \(f_I > f_{\text{rep}}/4\), the frequency \(f_Q\) exceeds \(f_{\text{rep}}/2\), causing Channel Q to undergo spectral aliasing because the low-pass filter (LPF) selects a different comb-line pairing. This shifts the comb-tooth index pairing between Channel I and Channel Q by 1, introducing a time delay offset of \(2\pi \beta_2 f_{\text{r,A}}\). Consequently, a step-like discontinuity appears in the \(\tau_{IQ}\) curve when \(f_I > f_{\text{rep}}/4\). Similarly, this higher-frequency region also exhibits central anti-symmetry, but centered about \(f_I = f_{\text{rep}}/2 - f_{\text{AOM}}/2\).

\begin{figure}
    \centering
    \includegraphics[width=\linewidth]{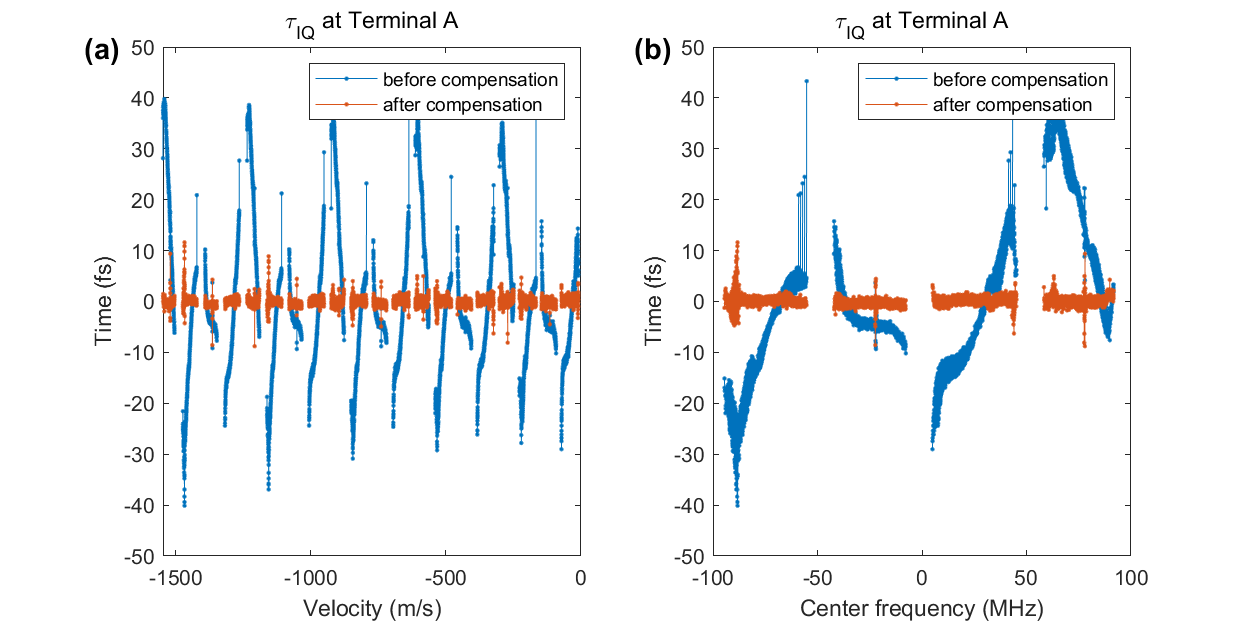}
    \caption{\textbf{Time difference between I and Q channels (\(\tau_{\text{IQ}}\)) before and after compensation.} \textbf{a}, \(\tau_{\text{IQ}}\) as a function of the simulated radial velocity. The uncompensated curve (blue) exhibits large periodic oscillations arising from the nonlinear RF response and comb-tooth misalignment. After applying the joint correction (orange), the residual errors are significantly suppressed over the full velocity range. \textbf{b}, \(\tau_{\text{IQ}}\) as a function of the center RF frequency of the I-channel interferogram. The uncompensated data (blue) reveal the continuous antisymmetric signature of the RF nonlinear phase, interrupted by discrete stepwise offsets induced by tooth-index slips due to comb chromatic dispersion. Both effects are completely corrected, leaving residual timing errors within \(\pm 2\) fs.}
    \label{fig:IQ_delay}
\end{figure}

Because the nonlinear RF response and chromatic dispersion reflected in the \(\tau_{IQ}\) curve depend strictly on the radio-frequency spectrum, \(\tau_{IQ}(v)\) exhibits a velocity-dependent periodicity given by \(v_{\text{period}} = c \cdot f_{\text{rep}} / f_{\text{CW}}\). The experimental results shown in Fig.~\ref{fig:IQ_delay} agree well with our theoretical model. The slight deviations from exact symmetry and the weak velocity dependence observed in the measured \(\tau_{IQ}\) curves (Fig.~\ref{fig:IQ_delay}a) are primarily attributed to the constant repetition rate approximation assumed in the theoretical derivation.

The systematic offset induced by optical dispersion can be conveniently read from the step height in the \(\tau_{IQ}\) plot. Processing the nonlinear RF response is more complex. Assuming \(\Phi(f)\) represents the RF phase response introduced by the detection system for each discrete FFT frequency bin \(f_i = i \cdot f_{\text{bin}}\), there are a total of \(\frac{N_s}{2}+1\) variables to solve, where \(N_s\) is the sample length of a single interferogram. Each measured value of $\tau_{IQ}(f)$ corresponds to an equation that involves $\Phi(f_i)$ for $i = 1, \ldots, \frac{N_s}{2}+1$. By acquiring \(\tau_{IQ}\) over a full Doppler sweep, we construct an overdetermined linear system and solve for \(\Phi(f_i)\) using linear programming or least-squares fitting. From this solution, the correction term \(\tau_{\text{RF}}\) for any velocity is computed. The choice \(f_{\text{AOM}} \neq f_{\text{rep}}/4\) avoids phase ambiguity singularities at exactly half the repetition frequency (see Supplemental Material for details). After applying both the RF response correction and the optical dispersion correction, the residual errors in the \(\tau_{IQ}\) plot are reduced to within \(\pm 2\text{ fs}\), see Fig.~\ref{fig:IQ_delay}b.

\section{Data availability}
The data supporting the plots within this article and other findings of this study are available from the corresponding author upon reasonable request.

\section{Code availability}
All relevant codes or algorithms are available from the corresponding author upon reasonable request.

\section*{References}

% --- bibliography inlined from sn-article-V5.bbl (no bibtex needed) ---

\section*{Acknowledgements}
This work is supported by the National Natural Science Foundation of China(grant no. T2125010, 61825505, and 123B2061); Quantum Science and Technology-National Science and Technology Major Project (Grant no. 2021ZD0300105, 2021ZD0300903, and 2021ZD0300904);Anhui Initiative in Quantum Information Technologies (grant no. AHY010100); Shanghai Municipal Science and Technology Major Project (grant no. 2019SHZDZX01); Shandong Postdoctoral Science Foundation.

\section*{Author contributions}
J.-W.P., Q.Z., and H.-F.J. conceived the project, supervised the research, designed the experimental schemes, and finalized the manuscript. 
J.-Y.G. carried out the theoretical analysis, conducted the laboratory technology verifications, participated in the field experiments, processed the data, and drafted the manuscript. 
Q.S. developed the low-phase-shift detection system, participated in the laboratory validations and field experiments, and contributed to writing the manuscript. 
L.H. developed and maintained the high-power optical frequency comb system, participated in the laboratory tests and field experiments, and contributed to writing the manuscript. 
J.-J.H. participated in the field experiments (leading the Nanshan ground station), prepared the figures, and assisted in drafting the manuscript. 
S.Z. and D.-Y.Z. contributed to the development of the low-phase-shift photodetectors. 
F.-H.C. participated in the field experiments and assisted in preparing the figures. 
Y.-C.Z., S.C., Y.-D.H. (contributing to prototype restoration and operation), Z.-K.Q., and Y.W. (assisting with optical link alignment and tracking) carried out the field experiments. 
M.L. participated in the laboratory validation and developed the hardware and logic for 2-fold repetition-rate sampling. 
S.-M.W. conducted the laboratory testing of the compact ultrastable laser and provided remote technical guidance during the field trials. 
F.-X.C. and L.-B.L. developed the fiber-optic time-frequency transfer system used for out-of-loop verification. 
Aili and J.-G.R. coordinated the logistics and administrative support at the Nanshan station, with J.-G.R. also assisting in telescope tracking and free-space link establishment. 
J.Y., R.S., and C.-Z.P. participated in discussing the experimental schemes and overall plans. 
All authors discussed the results and reviewed the manuscript.

\section*{Competing interests}
The authors declare no competing interests.

\clearpage
\setcounter{secnumdepth}{3}
\setcounter{section}{0}
\setcounter{subsection}{0}
\setcounter{subsubsection}{0}
\setcounter{figure}{0}
\setcounter{table}{0}
\renewcommand{\thesection}{S\arabic{section}}
\renewcommand{\thefigure}{S\arabic{figure}}
\renewcommand{\thetable}{S\arabic{table}}

\begin{center}
{\Large\bfseries Supplementary Information}\\[0.6em]
{\large Femtosecond Free-Space Time-Frequency Transfer at km/s Relative Motion with Doppler Compensation}
\end{center}

\vspace{0.5em}

\section{Motion-Compensated Time-Frequency Transfer Algorithm}
\label{sec:doppler_algorithm}
%%%%%%%%%%%%%%%%%%%%%%%%%%%%%%%%%%%%%%%%%%%%%%%%%%%%%%%%%%%%%%%%%%%%%%%%%%%%%%%%

The complete time-frequency transfer algorithm under Doppler conditions consists of three main components: one-way data processing, post-processing compensation (of one-way data), and two-way data fusion.

\subsubsection*{One-way Data Processing}

The purpose of one-way data processing is to convert the interferogram data acquired by the digitizer into one-way arrival-time data. At the same time, the auxiliary data stored includes the acquisition time of each data frame, whether it belongs to I data or Q data, the current speed, and the current acceleration. The flow of the algorithm is as follows.

\begin{enumerate}
    \item Input an interferogram frame and its acquisition timestamp.
    \item Calculate the current velocity using the velocity iteration algorithm.
    \item Determine whether the current interferogram is I-channel data (I data) or Q-channel data (Q data) using the IQ data identification algorithm.
    \item Calculate the phase curve of the zero-velocity interferogram using the interferogram reconstruction algorithm.
    \item If this is the first frame of the I or Q data, set it as the reference frame with a relative arrival time \(T_{arr}=0\), and return to step 1. Otherwise, calculate the one-way arrival time using the static time-frequency transfer algorithm.
    \item Feed the arrival-time result into the velocity iteration algorithm to resolve ambiguity and update the parameters of the system motion model, such as current velocity and acceleration.
    \item Return to step 1.
\end{enumerate}

Compared with the static time-frequency transfer algorithm, the additional core components mainly include the IQ data identification algorithm, the interferogram reconstruction algorithm, and the velocity iteration calculation algorithm.

\subsection{IQ Data Identification Algorithm}
The purpose of the IQ data identification algorithm is to determine whether the arriving interferogram belongs to I data or Q data based on the current velocity and system parameters. We distinguish them according to the spectrum of the interferogram. Based on the repetition rate difference between the two combs at zero velocity ($\Delta f_{\text{rep}}=f_{\text{rep}}^{(S)}-f_{\text{rep}}^{(LO)}$), the CEO locking frequency ($f_{\text{CEO}}$), the center frequency of the comb spectrum ($f_{\text{carrier}}$), the corresponding tooth index ($N_{\text{carrier}}$) and the comb repetition rate ($f_{\text{rep}}^{(LO)}$), we can calculate the center frequency position of the interferogram at zero velocity as

\begin{equation}
    f_{\text{center}}(0)=\text{abs}\left[ \text{mod}\left(f_{\text{CEO}}^{(S)}-f_{\text{CEO}}^{(LO)}+N_{\text{carrier}}\Delta f_{\text{rep}}, f_{\text{rep}}^{(LO)}\right)\right].
\end{equation}

Here we suppose that the $mod$ function returns a value between $-f_{\text{rep}}/2$ and $f_{\text{rep}}/2$. Thus, the center frequency position of the interferogram at any velocity $v$ can be written as

\begin{align}
    f_{\text{center}}^{(I)}(v)&=\text{abs}\left[ \text{mod}\left(f_{\text{CEO}}^{(S)}-f_{\text{CEO}}^{(LO)}+N_{\text{carrier}}\Delta f_{\text{rep}}-\frac{v}{c}f_{\text{carrier}}, f_{\text{rep}}\right)\right],
    \\
    f_{\text{center}}^{(Q)}(v)&=\text{abs}\left[ \text{mod}\left(f_{\text{CEO}}^{(S)}-f_{\text{CEO}}^{(LO)}+N_{\text{carrier}}\Delta f_{\text{rep}}-\frac{v}{c}f_{\text{carrier}}-f_{\text{AOM}}, f_{\text{rep}}\right)\right],
\end{align}

By comparing the actual interferogram center spectral position with $f_{\text{center}}^{(I)}(v)$ and $f_{\text{center}}^{(Q)}(v)$, we can determine whether it is I data or Q data.

There is one scenario that can hinder the above IQ frame identification, namely when $f_{\text{center}}^{(I)}(v) = -f_{\text{center}}^{(Q)}(v)$. In this case, the I and Q data have almost identical spectral positions but opposite spectral shapes and dispersion directions. The sign of the dispersion can be used to distinguish I data from Q data. Only after correctly identifying whether it is I or Q data will the dispersion of the transformed phase curve be consistent with the reference frame, yielding a well-fitted straight line after subtraction. Otherwise, the residual dispersion will be amplified after the reference frame is subtracted, resulting in a residual phase that follows a quadratic curve.

\subsection{Interferogram Reconstruction Algorithm}

The core challenge in high-velocity time-frequency transfer is the motion-induced envelope distortion of interferograms. When the velocity $v$ is significant, both the carrier frequency and the repetition rate of the signal comb change:

\begin{align}
f_c' &= f_c\left(1 - \frac{v}{c}\right) \\
f_{\text{rep}}' &= f_{\text{rep}}\left(1 - \frac{v}{c}\right)
\end{align}

This produces two distinct effects: (1) the spectral shift proportional to the $f_{\text{carrier}}$ change and (2) the spectral stretching proportional to the $\Delta f_{\text{rep}}$ change. The interferogram recovery algorithm compensates for both effects through frequency-domain resampling.

Given an estimated velocity $v$, the transformation proceeds as follows:

\begin{enumerate}
    \item Calculate the tooth indices corresponding to each frequency bin within the effective frequency range after the FFT of the reference frame.
    \item Based on the current velocity and the IQ frame identification result, calculate the corresponding FFT indices (fractional) for these teeth in the current environment.
    \item Use interpolation (resampling) to compute the phases corresponding to these teeth.
\end{enumerate}

The calculated phases constitute the recovered phase. Subsequently, using the static time-frequency transfer algorithm, the recovered phase is subtracted from the corresponding reference-frame phase. Linear fitting is then performed on the phase difference, and the one-way arrival time is calculated from the slope.

The delay-Doppler coupling is inherently canceled in this approach because the transformation explicitly accounts for the velocity-dependent spectral modifications.

\subsection{Velocity Iteration Calculation}

The initially input velocity (e.g., obtained from the laser ranging system) carries certain errors, which cause inaccurate spectral position calculation in the interferogram recovery algorithm, thereby introducing arrival-time calculation errors. Therefore, we first assume a motion model for the system over second-level timescales, such as a uniformly accelerated model, and then use the current and past time-of-flight data to calculate the velocity and acceleration parameters required by this model. To reduce computational load, we simply use three data points within the past 1 second to calculate velocity and acceleration.

When calculating the arrival time, ambiguity resolution must be performed first, which requires knowledge of the approximate arrival time information. This can be obtained from the motion parameters (velocity and acceleration) of the previous moment. The calculated arrival time is then used to update the current motion parameters and to estimate the velocity and ambiguity numbers for the next interferogram frame.

\begin{figure}
    \centering
    \includegraphics[width=\linewidth]{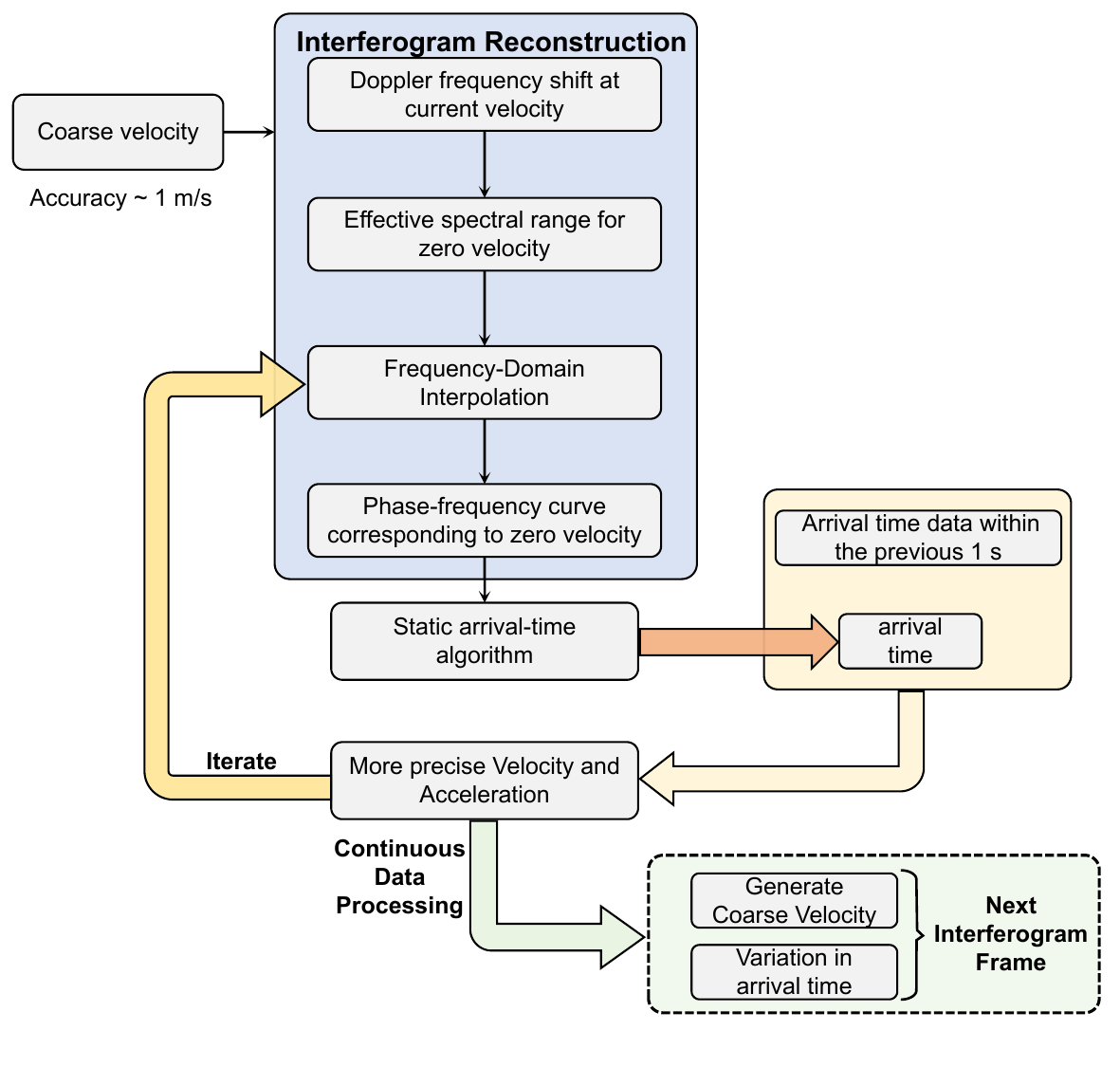}
    \caption{Main process of single-frame data processing. The main input is coarse velocity, which is used to calculate the precise one-way arrival time through Interferogram Reconstruction iterative algorithm, while also calculating velocity and acceleration as inputs for the next frame.}
    \label{fig:dataprocess}
\end{figure}

In general, this is a sequential iterative process, see Fig.~\ref{fig:dataprocess}. Only an initial coarse velocity input is required at the very beginning. Subsequently, as long as the link is not interrupted for an extended period, the system can sustain itself through high-precision velocity iteration. The specific procedure is as follows:

\begin{enumerate}
    \item If this is the first frame, use the external initial velocity and set the initial ambiguity number to 0. Otherwise, use the velocity and approximate arrival time calculated from the motion model.
    \item Feed the velocity information to the interferogram recovery algorithm and further calculate the time of flight.
    \item Perform an ambiguity resolution using the data from step 1 to obtain the correct time of flight.
    \item Update velocity and acceleration based on the current time of flight and two historical arrival time data points.
    \item Repeat step 2 using the updated parameters, iterating a specified number of times to obtain a sufficiently precise velocity and arrival time.
    \item Wait for the next frame input and return to step 1.
\end{enumerate}

Typically, convergence is achieved in 2 iterations starting from the GNSS velocity accuracy ($\sim$0.2 m/s), reaching cm/s precision sufficient for sub-femtosecond timing. A simple Kalman filter is used to eliminate the influence of occasional false triggers.

\subsubsection*{Post-Processing Compensation}

Post-processing of the data is a key step in eliminating errors introduced by the Doppler effect. Mostly, it includes removing errors caused by nonlinear RF response and optical dispersion. In addition, IQ data fusion is also performed at this stage.

\subsection{IQ Data Fusion}

The quadrant-shifted dual-channel linear optical sampling architecture generates two complementary data streams (Channels~I and Q). Because the local comb in Channel~Q is delayed by $\sim 1.5$\,ns and frequency-shifted by an AOM, each stream records a slightly shifted one-way arrival time. Performing robust IQ data fusion is essential to seamlessly bypass periodic Nyquist dead zones and maintain continuous timing measurements, while simultaneously providing the zero-centered differential timing observable required for subsequent calibration of the electronic RF nonlinearity and optical dispersion.

The procedure for IQ Data Fusion proceeds as follows:

\begin{enumerate}
    \item \textbf{Coincidence window selection}: Select valid IQ data pairs. To prevent extrapolation errors from degrading timing accuracy, a 5-ms coincidence time window is established; an I-frame and a Q-frame are identified as a valid pair only if their acquisition timestamps fall within this window.
    \item \textbf{Dynamic time alignment}: Extrapolate the Q-frame arrival times to the corresponding I-frame time points using the instantaneous velocity and acceleration model, compensating for the slight sampling asynchrony between the two channels.
    \item \textbf{Baseline delay calibration}: Compute the raw differential arrival time for each valid pair. Applying a $3\sigma$ filter to reject measurement outliers, the nominal inter-channel baseline delay $\tau_{\text{delay}}'$ is extracted as the statistical mean:
    \begin{equation}
        \tau_{\text{delay}}' = \langle \tau_{\text{Q}} - \tau_{\text{I}} \rangle_{\text{valid pairs}}.
    \end{equation}
    Because separate initial reference frames are selected for Channels~I and Q, $\tau_{\text{delay}}'$ accounts for both the physical optical path difference ($\sim 1.5$\,ns) and the relative reference offset between the two channels, serving as the static DC baseline.
    \item \textbf{Baseline de-trending and seamless fusion}: Subtract the nominal baseline delay from all Q-channel arrival times, yielding the calibrated Channel~Q series $\tau_{\text{Q}}^{\text{aligned}} = \tau_{\text{Q}} - \tau_{\text{delay}}'$. This operation centers the inter-channel delay around zero, producing the zero-centered differential observable $\tau_{\text{IQ}}(v) = \tau_{\text{Q}}^{\text{aligned}}(v) - \tau_{\text{I}}(v)$ used for subsequent RF and dispersion inversions. Whenever Channel~I falls into a Nyquist dead zone, the valid, baseline-aligned Channel~Q measurement is seamlessly substituted, ensuring continuous, dropout-free time transfer across the entire orbit.
\end{enumerate}

\subsection{Compensation for Error Caused by RF Nonlinear Response and Optical Dispersion}

The effect of RF nonlinear response and optical dispersion dominated the residual errors. The final step of post-processing for one-way data is to compensate for these errors. Information about these effects is already contained in $\tau_{\text{IQ}}$. Without them, the curve $\tau_{\text{IQ}}$ should be a constant value at all velocities. To resolve the error, we calculate $\tau_{\text{IQ}}(v)$ corresponding to different velocities $v$, thereby obtaining a $\tau_{\text{IQ}}(v)$ curve. The shape of the curve is related to the error.

For error compensation, there are a total of $N_s/2+3$ variables to be solved, including the RF response curve $\Phi(f_i)$ for $i=1, \ldots, N_s/2+1$, the time error per tooth mismatch caused by optical dispersion $\tau_{\text{mismatch}}$, and the overall offset of the $\tau_{\text{IQ}}$ curve $\tau_{\text{bias}}$, where $N_s$ is the number of sampling points per interferogram frame and $f_i=i \cdot f_{\text{bin}}=i \cdot f_{\text{rep}}/N_s$.

Since both the interferogram transformation and the time-of-flight calculation can be approximated as linear transformations with respect to the spectral phase, we can use linear programming to solve for these variables. We group all $\tau_{\text{Q}} - \tau_{\text{I}}$ data by velocity, with each group spanning 0.1 m/s, resulting in $N_v$ groups of $\tau_{\text{IQ}}(v_j)$ for $j=1, \ldots, N_v$. Of course, some results do not have values due to dead zones, which we set to 0. We then construct a $N_v \times (N_s/2+3)$ matrix $\mathbf{M}$, where $M(j,i)$ represents the contribution of the $i$-th variable to $\tau_{\text{IQ}}(v_j)$. Similarly, dead zone cases are set to 0, indicating that the corresponding variables do not participate in subsequent solving.

\begin{figure}
    \centering
    \includegraphics[width=\linewidth]{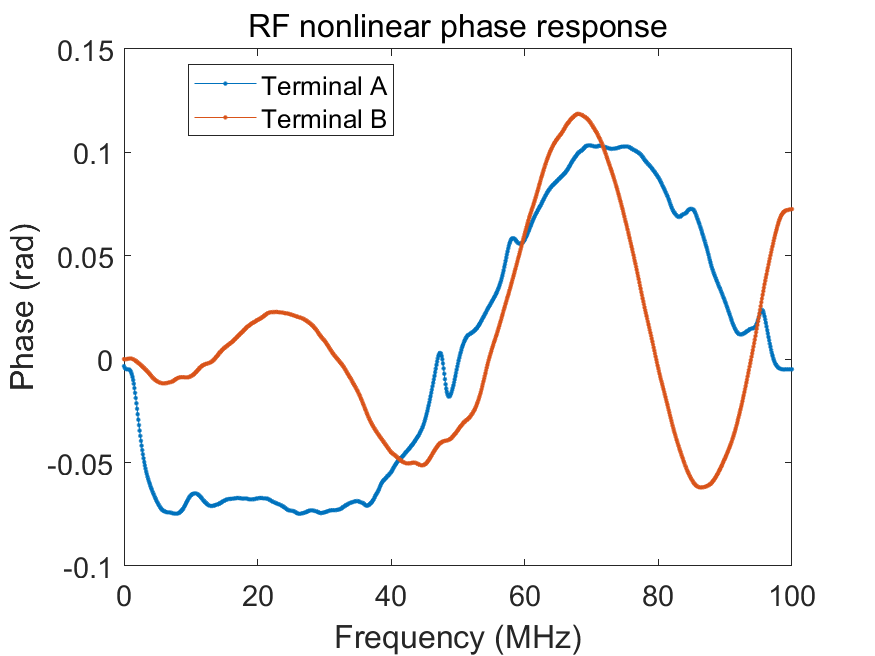}
    \caption{The solved RF nonlinear phase response of each Terminal.}
    \label{fig:S1}
\end{figure}

Finally, $\vec{x}=\text{pinv}(\mathbf{M}) \cdot \vec{\tau}_{\text{IQ}}$ gives the variables to be solved. Due to the influence of system statistical noise and drift, appropriate smoothing constraints must be added to obtain reasonable results for $\vec{x}$. The result of RF phase response is shown in Fig.~\ref{fig:S1}.

The residual error after compensation is typically $< \pm 2$~fs. The main sources of error are statistical noise and temperature drift between the I and Q paths. The latter has a long drift of about 5~fs in our system and can be further optimized later by designing a corresponding delay-locking system so that the delay between the I and Q paths is locked to a sub-femtosecond level using the reference CW.

After solving for the RF nonlinear response and optical dispersion, the delay compensation values for the I/Q arrival-time data at each velocity can be directly calculated, and subsequent data compensation can then be performed.

\subsubsection*{Dual-Terminal Data Fusion}

The final clock offset calculation combines data from both terminals. Due to Doppler effect, the up-link and down-link paths are asymmetric, introducing a velocity-dependent bias. This deviation is caused by two factors: asynchronous data acquisition at the two ends, and inconsistencies in round-trip flight time due to motion. We correct these two errors in turn.

To address the issue of asynchronous data acquisition, interpolation is required for the data from Terminal B, i.e., computing $T_{A\rightarrow B} (t_{B,m})$ as $T'_{A\rightarrow B}(t_{A,n})$ by extrapolation. The extrapolation is based on the velocity and acceleration information obtained during one-way data processing. Since we employ frequency shifting to simulate the Doppler effect, the interval of Terminal B's local clock varies with the frequency shift. If the actual sampling time were calculated directly as $\frac{n_{\rm timestamp}}{f_{\rm rep,B}}$ based on the timestamp count $n_{\rm timestamp}$, deviations would occur at the millisecond level or even higher. At high velocities, the bias introduced by such deviations becomes clearly visible. Here, based on the empirical observation that the variation of link delay $T_{\rm Link}$ is essentially less than 1~ns, we assume $\Delta T(t) \approx T_{A\rightarrow B}(t)-T_{A\rightarrow B}(0)$, thus we have

\begin{align}
    t_{B,m,real} \approx & t_{B,m} - T_{A\rightarrow B} (t_{B,m}),\\
    dt = & t_{A,n} - t_{B,m,real}, \\
    T'_{A\rightarrow B}(t_{A,n}) = & T_{A\rightarrow B} (t_{B,m}) + \frac{v}{c}dt+\frac{1}{2c}a\cdot dt^2.
\end{align}

Now we have both one-way data with the same timestamps. Next, we calculate the variations in the link and clock difference. Their relationship with the one-way arrival time can be written as

\begin{align}
    T'_{A\rightarrow B}(t)=&T_B(t)-T_A(t)-\frac{L(t)}{c},
    \\
    T_{B\rightarrow A}(t)=&T_A(t)-T_B(t-\frac{L(t)}{c})-\frac{L(t)}{c},
\end{align}
where $L(t_{recv})$ is the distance of the link at timestamp $t_{recv}$. According to the sign definition in our experiment, we have $T_B(t_{recv}-\frac{L(t_{recv})}{c}) = T_B(t_{recv})+\frac{v}{c}\frac{L(t_{recv})}{c}$. So we have

\begin{equation}
    \Delta T(t) = \frac{1}{2}\left[T'_{A\rightarrow B}(t)-T_{B\rightarrow A}(t)\right]-\frac{L \cdot v}{2c^2}.
\end{equation}

The residual systematic error $\frac{L \cdot v}{2c^2}\sim$1~ns in our experiment. The distance of the link $L$ and real-time velocity $v$ should be known to resolve the asymmetric error. $v$ can be read from one-way data. In our experiment, in order to obtain the value of L, we note that the link delay can be written as
\begin{equation}
    L(t) = -\frac{c}{2}\left[T'_{A\rightarrow B}(t)+T_{B\rightarrow A}(t)\right]-\frac{L \cdot v}{2c}
\end{equation}
The actual $L(t)$ should be a slow-varying function and independent of the frequency modulation velocity $v$. Based on this, one can fit the original result $\frac{c}{2}\left[T'_{A\rightarrow B}(t)+T_{B\rightarrow A}(t)\right]$ against v to obtain the value of L. Subsequently, the bidirectional asymmetric error $\frac{L \cdot v}{2c^2}$ can be eliminated, allowing for the calculation of the corrected clock difference $\Delta T(t)$.

\section{Detailed Experimental Setup}
\begin{figure}
    \centering
    \includegraphics[width=\linewidth]{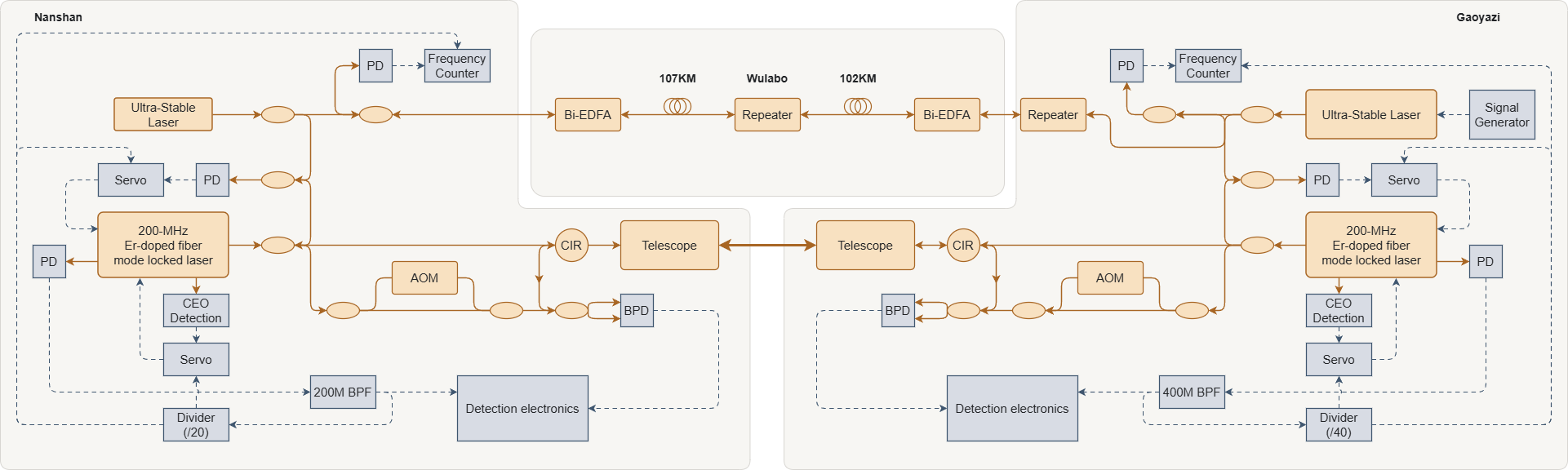}
    \caption{Schematic diagram of the experimental setup for the 113-km free-space time-frequency transfer. The system comprises two ground terminals located at Nanshan (Terminal A) and Gaoyazi (Terminal B). Both terminals are equipped with 200-MHz Er-doped fiber mode-locked lasers stabilized to ultrastable lasers via self-referenced locking schemes. The high-speed Doppler effect is emulated at Terminal B by dynamically tuning the ultrastable laser frequency through an offset-frequency locking technique with a signal generator and a portable cavity. The dual-channel sampling architecture, which includes acousto-optic modulators (AOMs) for the I and Q paths and specific band-pass filter divisions, is implemented to overcome Nyquist dead zones. A parallel fiber-optic link incorporating two bidirectional erbium-doped fiber amplifiers (Bi-EDFAs) and a Wulabo repeater provides the out-of-loop verification for the free-space link.}
    \label{fig:detail_setup}
\end{figure}

The experimental validation of the high-speed Doppler compensation scheme was conducted in Urumqi, Xinjiang Province. The system comprises two terminals located at Nanshan (Terminal A) and Gaoyazi (Terminal B). These stations are separated by a 113-kilometer free-space atmospheric link. Each Terminal's core equipment includes a high-power optical frequency comb, an ultra-stable laser, and a telescope. In addition, a parallel fiber-optic frequency transfer link serves as an out-of-loop verification system. The schematic diagram of the complete experimental apparatus is presented in Fig.~\ref{fig:detail_setup}.

\subsection{Frequency Combs and Ultrastable Lasers}

Each ground terminal is equipped with an Er-doped fiber mode-locked laser that generates an optical frequency comb. The comb operates at a fundamental repetition rate (\( f_{\text{rep}} \)) of about 200~MHz. The spectral power of the comb is approximately 1 W within the operational wavelength band from 1558~nm to 1567~nm. The output pulse width is broadened to around 80~ps through dispersion management to mitigate nonlinear effects induced by high peak power. The dispersion parameter of the comb system is approximately 11~ps/nm.

Both frequency combs are stabilized to a 1550.12-nm ultrastable laser using a self-referenced locking scheme. This configuration ensures that the carrier-envelope offset frequency (\( f_{\text{CEO}} \)) and the beat frequency between the comb and the continuous-wave laser are coherently linked to the repetition rate. The repetition rate difference (\( \Delta f_{\text{r}} \)) between the two combs at the Nanshan and Gaoyazi terminals is maintained at around 2 kHz.

To realize self-referenced locking, a small portion of the comb light is first directed to a photodetector to generate the repetition rate and its higher-order harmonics. At the Nanshan station, a 200 MHz band-pass filter isolates the fundamental repetition rate signal. After passing through a three-way power splitter, one channel is delivered to the digitizer while the other undergoes a 20-fold frequency division to obtain a 10 MHz signal for locking the CEO and beat frequency. The digitizer at the Nanshan station has a built-in phase-locked loop, which locks its internal 400 MHz clock to the input 200 MHz repetition rate and performs sampling at 400 MHz. At the Gaoyazi station, a 400 MHz band-pass filter extracts the second harmonic of the repetition rate. This signal is similarly split into two channels: one goes to the digitizer and the other is divided by 40 to yield a 10 MHz signal for the CEO and beat frequency locking. This is because the digitizer hardware at Gaoyazi does not have an internal PLL, so an external 400 MHz clock is required for its operation. At the Gaoyazi station, this self-referenced 10 MHz signal is also utilized as the reference clock for the frequency counter measuring the out-of-loop fiber beat note, thereby synchronizing the clock of the verification system with the local comb repetition rate.

\subsection{High-Speed Motion and Doppler Emulation Apparatus}

The Doppler effect associated with satellite motion is emulated at Terminal B using a broadly tunable ultrastable laser. This laser employs an offset-frequency Pound-Drever-Hall locking scheme. The ultrastable cavity is 2.5\,cm in length and is designed for portable transportation. A high-frequency signal generator drives an electro-optic modulator to introduce a variable frequency gap (\( f_{\text{EOM}} \)) between the seed laser and the stable optical cavity.

The frequency of the tunable ultrastable laser (\( f_{\text{CW,B}} \)) is governed by the relation \( f_{\text{CW,B}} = f_{\text{cavity}} + f_{\text{EOM}} \). The output frequency is dynamically adjusted by modulating \( f_{\text{EOM}} \) according to prescribed velocity profiles. This frequency shift on the continuous-wave reference laser is transferred to the entire optical frequency comb via the self-referenced locking mechanism. Consequently, the comb exhibits a synchronized breathing behavior that perfectly mimics the spectral stretching of a realistic relativistic Doppler shift. In this experiment, the maximum tuning range of \( f_{\text{EOM}} \) is from 500\,MHz to 1.5\,GHz. This range corresponds to an equivalent radial velocity of up to 1.55\,km/s for an optical carrier at 1550\,nm.

\subsection{Dual-Channel Linear Optical Sampling Detection}

The dual-channel linear optical sampling is implemented to mitigate Nyquist dead zones and retrieve the effects of both RF and optical dispersion. The local oscillator comb is divided into two optical paths. The first path serves as channel I without additional frequency shifting. The second path is routed through an acousto-optic modulator for channel Q. The AOM is driven at a radio frequency of approximately 47.26\,MHz. This frequency is intentionally set close to one quarter of the fundamental repetition rate. A fiber delay line of about 1.5\,ns is inserted into the Q path to separate the interferogram positions. The two channels are recombined with the incoming signal comb. A balanced photodetector detects the interference signals.

\subsection{Fiber-Optic Out-of-Loop Verification}

An independent fiber-optic frequency transfer link serves as the out-of-loop verification system. This link uses two bidirectional erbium-doped fiber amplifiers to compensate for propagation loss. A repeater unit is installed at the intermediate node of Wulabo. The repeater divides the total fiber link into two segments of 107 km and 102 km. This division significantly improves the noise suppression performance of the fiber frequency transfer link, especially when the fiber link is entirely aerial. The fiber-delivered signal from Terminal A is finally combined with the local ultrastable laser B at Terminal B. A photodetector records the resulting heterodyne beat note. An \(\Pi\)-type frequency counter measures the phase of this beat signal. The beat signal undergoes a 16-fold frequency division before being recorded because the maximum input range of the frequency counter is limited to 110\,MHz. The reference clock of this frequency counter is synchronized with the repetition rate of the optical comb at Terminal B. This synchronization ensures that the out-of-loop verification system shares a common time base with the free-space transfer system. The independent measurement provides a precise reference for evaluating the residual systematic errors. It validates the performance of the free-space time-frequency link under the simulated Doppler conditions.

\section{Design of Low Group-delay Distortion Detection System}
\begin{figure}
    \centering
    \includegraphics[width=\linewidth]{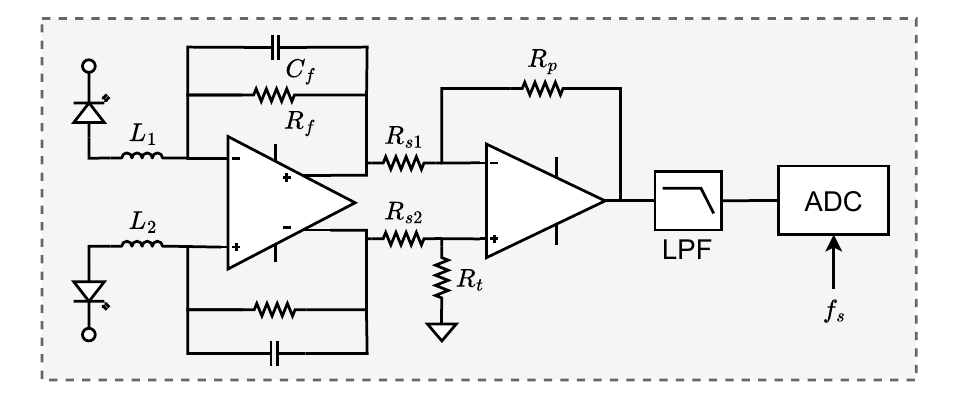}
    \caption{Structure of the low group-delay distortion detection system.}
    \label{fig:S2}
\end{figure}

To minimize the influence of the nonlinear phase of the linear optical sampling detection system on the time measurement accuracy with a high Doppler effect, we implemented a low group delay distortion detection system with a flat group delay response, also known as low phase-shift. The detection system includes a balanced detector, a low-pass filter, and an analog-to-digital converter (ADC) circuit, as shown in  Fig.~\ref{fig:S2}. The balanced detector is composed of two-stage amplifiers. The first-stage amplifier is a trans-impedance amplifier for photocurrent based on paired photodiodes, followed by a voltage amplifier as the second stage. To achieve a balance between detector noise performance, bandwidth, and group-delay flatness, high frequency inductors are inserted in the input stage to isolate the junction capacitance of the photodiode from the input capacitance of the amplifier. For our system, the effective RF frequency range is up to half the repetition rate of the optical frequency comb of 200 MHz. So the bandwidth of the detector should be larger than 100 MHz. Meanwhile, the detection system should have high frequency suppression performance to reduce high frequency harmonic aliasing and common-mode baseline voltage interference, especially at the optical comb repetition frequency. A Bessel-type low-pass filter is used here to maintain a flat group delay within the effective bandwidth. To further reduce high harmonic aliasing, a high speed ADC is employed, working with the sampling clock at twice the repetition rate of the optical comb.

%%%%%%%%%%%%%%%%%%%%%%%%%%%%%%%%%%%%%%%%%%%%%%%%%%%%%%%%%%%%%%%%%%%%%%%%%%%%%%%%

\end{document}